\documentclass{arxiv}
\usepackage{eurovis2026}

 \SpecialIssuePaper         

\CGFccby

\usepackage[T1]{fontenc}
\usepackage{dfadobe}  

\usepackage{cite}  
\BibtexOrBiblatex
\electronicVersion
\PrintedOrElectronic
\ifpdf \usepackage[pdftex]{graphicx} \pdfcompresslevel=9
\else \usepackage[dvips]{graphicx} \fi

\graphicspath{{./figures/}}

\usepackage{egweblnk}

\title{Noisy Graph Patterns via Ordered Matrices}

\author[J. Wulms, W. Meulemans, \& B. Speckmann]
{\parbox{\textwidth}{\centering J. Wulms\thanks{J. Wulms and W. Meulemans are (partially) supported by the Dutch Research Council (NWO) under project number VI.Vidi.223.137.}\orcid{0000-0002-9314-8260},
        W. Meulemans{$^\dagger$}\orcid{0000-0002-4978-3400},
        and
        B. Speckmann\orcid{0000-0002-8514-7858}
        }
         \\         
{\parbox{\textwidth}{\centering TU Eindhoven, the Netherlands}}%
}

\volume{36}   
\issue{1}     
\pStartPage{1}      

\usepackage{amsmath}
\usepackage{amssymb}

\usepackage{multirow}

\usepackage{wrapfig}

\usepackage{algorithm}
\usepackage[noend]{algorithmic}

\newcommand{\mypar}[1]{\smallskip\noindent\textbf{#1}}

\usepackage{hypcap}

\begin{document}

\teaser{
 \includegraphics[width=0.95\linewidth]{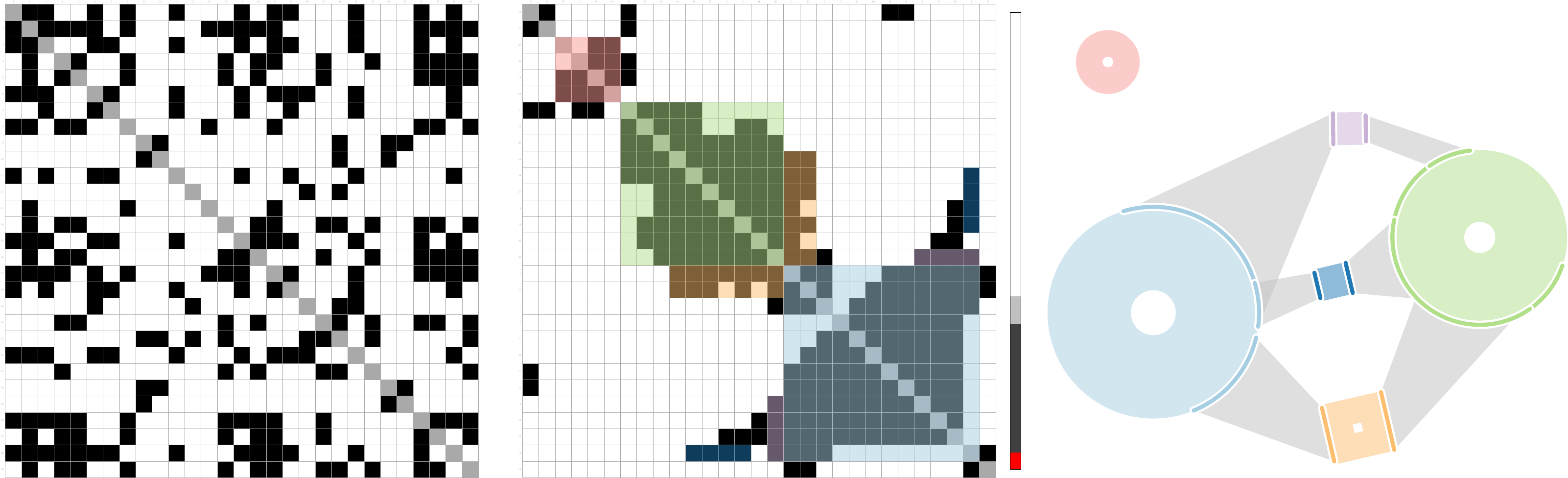}
 \centering
  \caption{Detecting and visualizing \emph{noisy patterns} in a graph. (Left) The input of our pipeline: a plain adjacency matrix~$M$. (Middle) The output of our pipeline: matrix~$M$ reordered to optimize Moran's $I$ such that it exhibits graph patterns as rectangles in the matrix. Our pipeline then detects noisy patterns with sufficient black-black adjacencies; (Right) Our \emph{Ring Motif} simplification: the high-level structure of the input graph can be summarized by visualizing the detected patterns as motif glyphs that explicitly encode the noise levels of the detected patterns.}
\label{fig:teaser}
}

\maketitle
\begin{abstract}
    The high-level structure of a graph is a crucial ingredient for the analysis and visualization of relational data. However, discovering the salient \emph{graph patterns} that form this structure is notoriously difficult for two reasons. (1) Finding important patterns, such as cliques and bicliques, is computationally hard. (2) Real-world graphs contain noise, and therefore do not always exhibit patterns in their pure form. Defining meaningful \emph{noisy patterns} and detecting them efficiently is a currently unsolved challenge.
    In this paper, we propose to use well-ordered matrices as a tool to both define and effectively detect noisy patterns. Specifically, we represent a graph as its adjacency matrix and optimally order it using Moran's $I$. Standard graph patterns (cliques, bicliques, and stars) now translate to rectangular submatrices. Using Moran's $I$, we define a permitted level of noise for such patterns. A combination of exact algorithms and heuristics allows us to efficiently decompose the matrix into noisy patterns.
    We also introduce a novel motif simplification that visualizes noisy patterns while explicitly encoding the level of noise. We showcase our techniques on several real-world data sets.


\end{abstract}  
\section{Introduction}

Graphs are a ubiquitous tool to model relational data: analyzing and visualizing graphs has myriad applications, for example, in biology, logistics, humanities, social interactions, and business intelligence.
The high-level structure of a graph is a crucial ingredient to support such analysis and visualization, and is typically considered to consist of salient \emph{graph patterns} and their interconnections. A common approach to visualize such patterns is via \emph{motif simplification}, replacing patterns (subgraphs) by a glyph indicating their structure.

However, discovering graph patterns is notoriously difficult for two reasons.
First, it is computationally hard to reliably detect the most salient patterns, even for very crisply defined structures such as cliques or bicliques \cite{DBLP:journals/jal/Hochbaum98, DBLP:journals/eor/WuH15}. This is due to the fact that there is an exponential number of potential (bi)cliques to check. It is likely that almost all graph patterns are computationally hard to detect~\cite{DBLP:journals/siamcomp/DalirrooyfardVW21}. 
Second, real-world data sets often contain noise, and therefore typically deviate from patterns in their pure form. For example, some edges may be missing to formally consider a set of vertices as a clique, yet their interconnectivity is sufficiently high to consider it a pattern for most intents and purposes. Defining meaningful \emph{noisy patterns} and detecting their most salient occurrences is a currently unsolved challenge.

\mypar{Contributions.}
We propose to leverage well-ordered matrices as a tool to both define and detect noisy patterns. Common patterns emerge as contiguous, rectangular submatrices, effectively reducing the potential candidate set from exponential (all subsets) to polynomial, as a submatrix is readily defined by its two opposite corners.
We use Moran's $I$, skewed to consider only adjacencies between present edges, to define the level of noise in a pattern in various ways.
In the remainder, we focus on cliques and bicliques: we show how to efficiently detect the optimal set of noisy cliques exactly, and introduce a heuristic to detect a good set of noisy bicliques.
Furthermore, we introduce a novel motif simplification that visualizes detected noisy patterns while explicitly encoding their level of noise. Our \emph{Ring Motif} simplification is a direct generalization of traditional node-link representations. Ring Motifs can be used as a stand-alone visualization of high-level graph structure captured by noisy patterns, or in combination with a matrix-view, which offers detailed view of individual edges. In this paper, we use the combined views with color-coordinated patterns.

After reviewing related work, we formally introduce our definitions of noisy patterns and the considerations therein in \autoref{sec:defining}. In \autoref{sec:detecting}, we describe our algorithms for efficiently detecting a good set of noisy patterns. Our visualization technique is described in \autoref{sec:visualizing}, and used to demonstrate the results of our techniques on several real-world data sets in \autoref{sec:case-study}.

\mypar{Related work.}
A plethora of techniques exists to visualize group structures in graphs~\cite{DBLP:journals/tvcg/JianuRHT14,DBLP:journals/cgf/VehlowBW17}. However, the groupings used by these techniques often do not represent the high-level graph patterns, but instead focus on visualizing an associated set system or other auxiliary data. When the grouping does represent the graph structure, the patterns are often predetermined, for example when the graph comes with a predefined hierarchy~\cite{DBLP:journals/tvcg/ArchambaultMA11,DBLP:journals/tvcg/WuNV22}. 

Among the techniques that do summary graph structures, Power Graph Decompositions~\cite{DBLP:journals/tvcg/DwyerRMM13,DBLP:conf/apvis/DwyerMMNMW14} stand out as a lossless compression technique to draw dense graphs with fewer edges. However, their intricate design makes optimal decompositions computationally hard to obtain. If some abstraction is allowed, we can use motif simplification~\cite{DBLP:conf/chi/DunneS13} to simplify a node-link diagram with glyphs summarizing cliques, fans, and certain bicliques. Recently, these motif simplifications have been adapted to different graph structures~\cite{DBLP:journals/tvcg/ZhouLSCCWW25} and extended to other types of network visualization~\cite{DBLP:conf/vis/FuchsD25}. However, all these techniques can find patterns only in their pure form. Additionally, the motif simplifications consider an exponential number of patterns, such as cliques, or resort to community detection heuristics~\cite{blondel2024fast} to find graph patterns. In contrast, our pipeline can detect noisy patterns and reduces the number of candidate graph patterns to a polynomial number, by utilizing well-ordered matrices.

To visually analyze or communicate about a graph, drawing its adjacency matrix is a common technique. In a well-ordered matrix, the vertices are ordered such that patterns emerge~\cite{DBLP:conf/apvis/MuellerML07a, DBLP:journals/cgf/BehrischBRSF16}. Recently, it was proposed to interpret the ordered matrix as a map and use spatial auto-correlation measures, in particular, Moran's $I$, to model ordering quality~\cite{DBLP:journals/tvcg/BeusekomMS22}. Effectively, it promotes a general sense of structure in the presence and absence of edges.
In particular, computing a matrix ordering that reveals salient patterns reduces to a traveling-salesperson problem: while this problem is generally hard, it is well studied for various heuristics~\cite{croes1958method}, but also allows solutions that are efficient in practice for moderately sized instances \cite{concorde,neos}.

\section{Defining Noisy Patterns}\label{sec:defining}

In the remainder of this paper, we assume we are given an unweighted, undirected graph $G=(V,E)$, where $E \subseteq {V \choose 2}$ without self-loops. That is, an edge is a set of two vertices $\{ u,v \} = \{ v,u \}$ with $u \neq v$. We assume that the graph has $n = |V|$ vertices, and $m = |E|$ edges.
While our definitions straightforwardly generalize to other types of graphs, investigating the efficacy of such generalizations is beyond the scope of this paper.

We introduce three guidelines that a good definition for noisy patterns in $G$ should adhere to.
\begin{description}
    \item[Dense:] Patterns in their pure form are typically defined by the presence of a set of edges. Noisy patterns should match this pure form well, even if edges are missing.
    \item[Structured:] Patterns reflect a certain structure of edges on a set of vertices. Noisy patterns should therefore be structured as well.
    \item[Tight:] As graphs vary in size and levels of noise, and different use cases may require different levels of detail of the graph structure, it is expected for noisy-pattern definitions to be parametrized. Yet, such parameters may introduce ``slack'': a low-noise pattern can be extended with unrelated vertices just to increase its size. We must be careful to not needlessly extend patterns this way when searching for large patterns.
\end{description}

Consider some bijective function $\rho \colon \{1,\ldots,n\} \rightarrow V$, an \emph{ordering} of the vertices. The adjacency matrix $M_\rho$ of $G$ is a symmetric binary matrix, with rows and columns matching vertices according to $\rho$. A cell $M_\rho[i,j]$ at row $i$ and column $j$ has value $1$ if $\{ \rho(i), \rho(j) \} \in E$, and value $0$ otherwise. We generally refer to and visualize 0-valued cells as \emph{white} and 1-valued cells as \emph{black}.

We observe that, generally, a pattern is defined by the existence of edges within or between one or more sets of vertices. For example, a clique is a set $C \subseteq V$ of vertices, such that all edges between vertices in $C$ exist; a biclique consists of two disjoint sets $A, B \subset V$, such that every edge from one set to the other exists.
In a well-ordered matrix $M_\rho$, the ordering makes such patterns visually emerge: cliques form black squares along the diagonal, and bicliques are black rectangles on one side of the diagonal. In either case, these visual shapes correspond to contiguous submatrices of~$M_\rho$. We use $M_\rho[i:i'\times j:j']$ to denote the submatrix spanning rows $i$ through $i'$ (inclusive) and columns $j$ through $j'$ (inclusive). We simplify notation to $M_\rho[i \times j:j']$ or $M_\rho[i:i' \times j]$ for a submatrix of a single or column, respectively.

To account for noisy patterns, we model how much such a submatrix deviates from the pattern in its pure form. We discuss several options below and how they match our general guidelines.

\mypar{Density.}
The perhaps most obvious choice is to compute the relative number of edges -- the number of black cells in the submatrix, divided by the total number of cells that should be black in the pure pattern. If this fraction is above a certain threshold, we consider the submatrix a noisy pattern. 
While this adheres to our \emph{Dense} guideline, it does not satisfy the \emph{Structured} and \emph{Tight} guidelines: a checkerboard has about half the number of edges, but represents two wholly disjoint subgraphs; and a perfectly black submatrix can easily be extended with a white row if the threshold allows for it.

\mypar{Moran's $I$.}
Given the power of reordering matrices based on Moran's $I$ \cite{DBLP:journals/tvcg/BeusekomMS22}, one could consider a definition that evaluates a submatrix based on its spatial autocorrelation. However, while this is certainly \emph{Structured}, observe that a fully white submatrix is also (very) well correlated, and would thus be considered a pattern. Clearly, such cases do no satisfy the \emph{Dense} guideline. Moreover, the \emph{Tight} guideline is violated: adding empty rows or columns may in fact increase Moran's $I$.

\mypar{Globally reweighted Moran's $I$.}
We may adhere to the \emph{Dense} guideline by reweighting Moran's $I$, to effectively only measure the relative number of black-black adjacencies in the submatrix. This change ensures that a submatrix that scores well must be dense as well as structured. However, it still does not quite adhere to the \emph{Tight} guideline, though at least the measure cannot increase by adding very sparse rows or columns.

\mypar{Locally reweighted Moran's $I$.}
To also adhere to the \emph{Tight} guideline, we further modify the above measure. Specifically, we introduce two thresholds, $\sigma$ and $\tau$, to control the level of structure and tightness.
We require that the fraction of adjacent rows, that have a relative number of black-black adjacencies exceeding $\sigma$, is at least~$\tau$; we analogously constrain the adjacent columns.

\begin{figure*}
    \centering
    \includegraphics{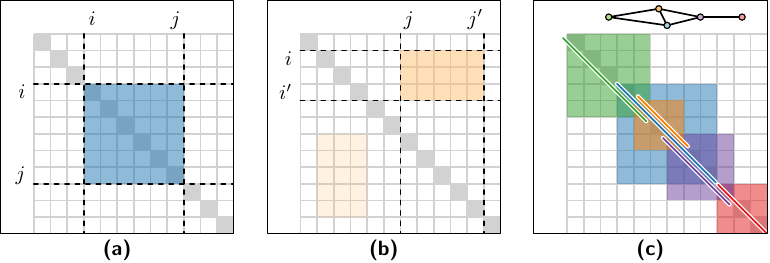}
    \caption{Elements of our noisy pattern detection pipeline. Defining and enumerating \textbf{\textsf{(a)}} cliques and \textbf{\textsf{(b)}} bicliques. \textbf{\textsf{(c)}} Selecting cliques.}
    \label{fig:pipeline}
\end{figure*}

\mypar{Specific noisy patterns.}
In our work, we focus on three patterns: 
\begin{description}
    \item[Clique:] a clique is defined by a single set of vertices, requiring all edges to be present between these vertices in its pure form. In a well-ordered matrix, it emerges as a black square submatrix $M_\rho[i:j \times i:j]$ along the diagonal, defined by its first index $i$ and last index $j > i$ in ordering $\rho$ (see Figure~\ref{fig:pipeline}a). In our noisy-pattern definitions, we exclude the diagonal (and its adjacencies) from consideration, as we assume graphs do not have self-loops: this diagonal and its adjacencies carry no structural information.
    \item[Biclique:] a biclique is defined by two sets of vertices, requiring all edges from one set to the other to be present in its pure form. In a well-ordered matrix, it emerges as a black (not necessarily square) submatrix $M_\rho[i:i' \times j:j']$, fully above the diagonal: $i \leq i' < j \leq j'$ (see Figure~\ref{fig:pipeline}b). It is defined by two intervals: one interval of rows $i:i'$ and one of columns $j:j'$. Note that we assume the graph is undirected, and thus matrices are symmetric: a biclique $M_\rho[i:i' \times j:j']$ above the diagonal also occurs as $M_\rho[j:j' \times i:i']$ below the diagonal.
    \item[Star:] a star, sometimes also referred to as fan, represents a high-degree, well-connected vertex. Effectively, it is a special type of biclique where one of the two defining sets is in fact a singleton. As such, our noisy-star definition simply follows that of a biclique: however, for the locally reweighted Moran's $I$, there are no adjacent rows or columns (depending on orientation), and this is therefore not constrained. 
\end{description}

\mypar{Noisy-pattern decomposition.}
To explain the high-level structure of a graph, we want to decompose it into a set of noisy patterns.
Our goal is threefold: (1) patterns should explain the high-level structure in a concise manner, without leaving clear structures unexplained; (2) distinct patterns should not explain the same parts of the high-level structure; (3) patterns that are close to their pure form are preferred over more noisy ones.

We hence define a noisy-pattern decomposition, as a \emph{maximal} set of \emph{disjoint} patterns, such that each pattern adheres to the chosen noisy-pattern definition, such as locally reweighted Moran's $I$. 
Two patterns are disjoint if their submatrices are; the set is maximal if adding any other noisy pattern causes overlapping submatrices.

To measure the quality of the decomposition, we assign to each pattern a weight, its number of black-black adjacencies, and use the total weight of the patterns in the decomposition. This weight readily promotes the use of dense, structured patterns, and prefers larger patterns over smaller ones, for a concise explanation.

\section{Detecting Noisy Patterns}\label{sec:detecting}

We propose a pipeline that combines exact algorithms with heuristics to decompose a graph $G$, given as adjacency matrix $M$, into noisy patterns. While our pipeline can handle different noisy-pattern definitions, we describe the pipeline for locally reweighted Moran's $I$, with parameters~$\sigma$ and~$\tau$ as user parameters. 

Our pipeline has three steps, elaborated below. 
(1) Reorder input matrix $M$, optimizing for Moran's $I$. The reordered matrix~$M_\rho$ has sufficient structure to reliably find salient graph patterns. 
(2)~Enumerate candidate noisy patterns~$P^*$ using the~$\sigma$ and~$\tau$ parameters. 
(3) Select a pairwise disjoint, maximal set~$P\subseteq P^*$ of high weight.

\subsection{Matrix Reordering}\label{sec:alg-order}
The core idea is that a well-ordered matrix reveals salient graph patterns. We hence want to maximize Moran's~$I$ via matrix reordering. A plethora of algorithms and heuristics exists for this purpose; refer to \cite{DBLP:journals/tvcg/BeusekomMS22} for a summary of algorithms and quality criteria.

We use that optimizing a matrix ordering for an adjacency measure, such as Moran's $I$, is effectively finding a shortest traveling salesperson path (TSP)~\cite{lenstra1975some}. While the TSP problem is NP-hard, dedicated libraries, such as Concorde \cite{concorde}, have been developed to efficiently compute an optimal \emph{TSP tour} for moderately sized instances.
A TSP tour differs slightly from a TSP path; a tour returns to the starting vertex. We hence have to adapt our input matrix $M$ to obtain the intended TSP path that corresponds to the matrix ordering that optimizes Moran's $I$. 

Below, we explain how to model optimizing Moran's $I$ as a TSP path instance, and subsequently how to extract an optimal solution from a modified TSP tour instance.

\mypar{Reducing to TSP path.} 
Input matrix $M$ has $n$ rows and columns and $m\leq n^2$ black cells. 
To compute a matrix ordering~$\rho$ that maximizes Moran's $I$ via TSP, we need to define distances between the rows of $M$. 
The general Moran's $I$ formula can be reduced to a simplified form for 0-1 matrices; see \cite{DBLP:journals/tvcg/BeusekomMS22}:
\begin{align*}
    I &= B \cdot \frac{n}{2(n-1)} + W \cdot \frac{n}{2(n-1)(n^2-m) - 1} = c_B \cdot B + c_W \cdot W - 1.
\end{align*}
Here, $B$ and $W$ denote the number of black-black and white-white adjacencies in $M$, respectively. This simplified form naturally translates to a similarity measure~$s(M, u, v)$ between rows $u$ and $v$ of $M$.
\begin{align*}
    s(M, u, v) = c_B \cdot B^{\textbf{|}}(M, u, v) + c_W \cdot W^{\textbf{|}}(M, u, v)
\end{align*}
$B^{\textbf{|}}(M, u, v)$ and $W^{\textbf{|}}(M, u, v)$ denote the number of vertical black-black and white-white adjacencies, respectively, in case $u$ and $v$ are made adjacent in~$\rho$. To obtain the distance~$\delta_I(M, u, v)$ between rows $u$ and $v$, we take the inverse of the similarity $s(M, u, v)$.
\begin{align*}
    \delta_I(M, u, v) = 1 - s(M, u, v)
\end{align*}

Maximizing Moran's $I$ matches finding an ordering such that $\sum \delta_I(M, u,v)$ for all adjacent pairs $(u,v)$ is minimal. Hence, we define a complete, weighted graph $T = (V, V^2)$ on the vertices $V$ of $G$, with edge weights according to $\delta_I$. The optimal TSP path of $T$ is a vertex ordering that optimizes Moran's $I$ in the adjacency matrix.

\mypar{Using a TSP tour.} 
Concorde \cite{concorde} computes a TSP tour instead. Hence, we create $T'$ by adding a virtual vertex $\omega$ to~$T$, and set its distance to any other vertex to $0$. By definition, a shortest TSP tour on $T'$ consists of a shortest TSP path in $T$, and a connection between the first and last vertex of this path via $\omega$. Hence, we easily derive the optimal TSP path on $T$ from this TSP tour on $T'$.

  

  

\subsection{Pattern Enumeration}\label{sec:alg-enum-pattern}

We now enumerate a set $P^*$ of candidate patterns: the submatrices in $M_\rho$ adhering to the given noisy-patterns definition.
We distinguish between candidate cliques $P_C$, candidate bicliques~$P_B$, and candidate stars $P_S$ in~$M_\rho$, such that $P^* = P_C \cup P_B \cup P_S$. Our candidate enumeration differs per pattern type (see Algorithm~\ref{alg:pattern-enum}): for cliques we enumerate all possible patterns, but for bicliques and stars this becomes computationally infeasible in large graphs. Instead, we focus on enumerating maximal biclique and star patterns with high weights, leading to a smaller set of candidates.

Since we focus on noisy patterns defined via locally reweighed Moran's $I$ in our pipeline exposition, we introduce a few additional definitions to describe our pattern enumeration. Consider some submatrix $M' \subseteq M_\rho$. As before, we denote with $B^{\textbf{--}}(M', u, v)$ the number of (horizontal) black-black adjacencies between columns $u$ and $v$ within $M'$: the number of shared neighbors of $u$ and $v$ within the vertex set(s) of $M'$.
Analogously, $B^{\textbf{|}}(M', u, v)$ denotes the number of (vertical) black-black adjacencies between rows.

\mypar{Cliques.} 
An interval of row indices matches to a set of consecutive vertices that is potentially a noisy clique. We consider all $O(n^2)$ such intervals of length at least $3$, that is, all intervals $[i,\ldots,j]$ with $j \geq i+2$. 
For each such interval, we test whether the submatrix $M_\rho[i:i \times j:j]$ (see Figure~\ref{fig:pipeline}a) adheres to the noisy-pattern definition, and if so, we add it as a pattern to $P_C$.

Specifically, let $M'= M_\rho[i:i \times j:j]$ be the square submatrix along the diagonal of $M_\rho$ matching some interval $[i,\ldots,j]$ . For each row $u$ with $i \leq u < j$, we check whether it has sufficiently many black-black adjacencies with the next row: $B^{\textbf{|}}(M_C, u, u+1) > \sigma \cdot (j-i+1)$. If the number of rows for which this holds, is at least $\tau \cdot (j-i)$, the interval is a noisy clique according the the locally reweighted Moran's $I$ model.
Note: as $M'$ is a symmetric submatrix, it is sufficient to check only vertical black-black adjacencies.

\mypar{Bicliques.} 
For bicliques and stars, we need to consider only patterns above the diagonal of~$M_\rho$, as the matrix is symmetric. There are $O(n^4)$ potential candidate bicliques, defined by the $O(n)$ options for the the first and last row, and the first and last column of the submatrix (see Figure~\ref{fig:pipeline}b). We do not consider each of these patterns as candidates, since this leads to an infeasible number of candidates to consider, for anything but trivially sized graphs. 

Instead, for each pair $i,j$ with $i < j + 1 \leq n$, we attempt to find a submatrix representing a noisy biclique as follows: first, we test whether $M_\rho[i:i+1 \times j:j+1]$ is a noisy biclique -- if it is, we proceed to greedily extend this submatrix while keeping as invariant that the submatrix is a noisy biclique. Concretely, we set $i' := i+1$ and $j' := j+1$, such that $M' = M_\rho[i:i' \times j:j']$ is a noisy biclique.
If adding both a row and a column to $M'$ maintains a noisy biclique, we increase both $i'$ and $j'$ by $1$;
otherwise, if adding a column maintains a noisy biclique, we increase only $i'$ by $1$;
otherwise, if adding a row maintains a noisy biclique, we increase only $j'$ by $1$;
otherwise, we cannot further extend $M'$ and add it as a pattern to $P_B$.

In the locally reweighted Moran's $I$ model, we test whether $M' = M_\rho[i:i' \times j:j']$ is a noisy biclique for $\sigma$ and $\tau$ as follows: the number of pairs of consecutive rows $u$ and $v$, for which $B^{\textbf{|}}(p_B, u, v) > \sigma \cdot (j'-j+1)$ should be at least $\tau \cdot (i' - i)$ and analogously we check for the number consecutive pairs of columns $u'$ and $v'$ with sufficient $B^{\textbf{--}}(p_B, u', v')$.

This greedy procedure misses some candidate bicliques, when both the rows and columns of $M_B$ are grown, or when the number of rows is increased instead of the number of columns. To significantly decrease the number of candidate bicliques (from $O(n^4)$ down to $O(n^2)$) such choices are necessary. As we focus on finding maximal patterns, this greedy procedure aims to maximize pattern weight: it chooses an additional row and column, over just one of the two, as that is locally the best choice to increase weight.

\mypar{Stars.} 
Candidate enumeration for stars is largely the same as for bicliques. However, stars consist of only a single column or a single row, and hence a potential pattern $M' = M_\rho[i:i' \times j:j']$ has the requirement that $i=i'$ or $j=j'$. We consider each cell $M_\rho[i,j]$ with $i<j$ as the starting point of both a vertical and horizontal star pattern, $M^{\textbf{|}} = M_\rho[i:i' \times j]$ and $M^{\textbf{--}} = M_\rho[i \times j:j']$. For both, we test extensions only by adding a row or adding a column, respectively. The maximal extension that still satisfies the noisy-star definition is added to $P_S$. This is again a heuristic greedy procedure that adds up to $O(n^2)$ of the $O(n^3)$ possible stars.

\begin{algorithm}[t]
  \caption{\textsc{EnumerateNoisyPatterns}$(M_\rho, \sigma, \tau)$}\label{alg:pattern-enum}
  \begin{algorithmic}[1]
  \REQUIRE A well-ordered adjacency matrix $M_\rho$ and noisy pattern parameters $\sigma$ and $\tau$
  \ENSURE Noisy pattern candidate set~$P^*$
  
  \smallskip 
\COMMENT{Enumerate candidate cliques}
\FOR{each pair $i,j$ with $1 \leq i$ and $i + 2 \leq j < n$}
        \IF{$M_\rho[i:i\times j:j]$ is noisy clique for $\sigma$ and $\tau$}
            \STATE Add $M_\rho[i:i\times j:j]$ to $P_C$
        \ENDIF
\ENDFOR
\COMMENT{Enumerate candidate bicliques and stars}
\FOR{each pair $i,j$ with $1 \leq i < j < n$}
        \IF{$M_\rho[i:i+1 \times j:j+1]$ is a noisy biclique for $\sigma$ and $\tau$}
            \STATE $i'= i+1; j'= j+1$
            \STATE Increase $i'$ and/or $j'$ while this maintains a noisy biclique%
            \STATE Add $M_\rho[i:i' \times j:j']$ to $P_B$
        \ENDIF
        \smallskip
        \IF{$M_\rho[i \times j:j+4]$ is noisy star for $\sigma$ and $\tau$}
            \STATE $j'= j+4$
            \STATE Increase $j'$ while this maintains a noisy star
            \STATE Add $M_\rho[i \times j:j']$ to $P_S$
        \ENDIF
        \smallskip
        \IF{$M_\rho[i:i+4 \times j]$ is noisy star for $\sigma$ and $\tau$}
            \STATE $i'= i+4$
            \STATE Increase $i'$ while this maintains a noisy star
            \STATE Add $M_\rho[i:i' \times j]$ to $P_S$
        \ENDIF
\ENDFOR
        
\RETURN $P^* = P_C \cup P_B \cup P_S$
\end{algorithmic}
\end{algorithm}

\mypar{Efficiency.}
The straightforward implementation takes $O(n^4)$ time for cliques and $O(n^5)$ time for bicliques: testing a noisy-pattern definition on a submatrix is linear in its number of cells.
However, we can easily precompute, for all $i,j$, the number of vertical black-black adjacencies in $M_\rho[1:i \times 1:j]$ in $O(n^2)$ time, by traversing the matrix once; and analogously for the horizontal adjacencies. 
For the density and globally reweighted Moran's $I$ models, this information allows us to test whether a submatrix adheres to its pattern definition in $O(1)$ time. As such, we enumerate all cliques in $O(n^2)$ time and all bicliques and stars in $O(n^3)$ time.

For the locally reweighted Moran's $I$ definitions, this information allows us to compute $B^{\textbf{|}}(M', u, u+1)$ and $B^{\textbf{--}}(M', u, u+1)$ in constant time for some submatrix $M'$. Testing whether a row or column within a submatrix satisfies $\sigma$ takes constant time, and testing whether their relative fraction exceeds $\tau$ is linear in the number of rows and columns. Thus, in this model, we enumerate all cliques and stars in $O(n^3)$ time and all bicliques in $O(n^4)$ time.

\subsection{Pattern Selection}\label{sec:alg-select-pattern}

Finally, we select a set of patterns from the candidate pattern set~$P^*$ computed in the previous step. The selected patterns should be disjoint and have high weight. We again distinguish between pattern types (see Algorithm~\ref{alg:pattern-select}): we first select an optimal set of clique patterns in~$P_C$ and then heuristically add biclique and star patterns from $P_B \cup P_S$. 

\mypar{Selecting cliques.} 
Recall that each noisy-clique pattern is defined by an interval $[i,\ldots, j]$. 
We define the following interval graph (see Figure~\ref{fig:pipeline}c): each vertex is a noisy-clique pattern in $P_C$, weighted according to the weight of the pattern; two vertices share an edge if the intervals defining their patterns overlap.
A disjoint set of patterns in~$P_C$ of maximum weight corresponds to a maximum weight independent set in the described interval graph. We can therefore use an existing (exact) algorithm to compute a \emph{maximum weighted independent set} (MWIS) in interval graphs~\cite{DBLP:journals/ipl/HsiaoTC92} to find an optimal disjoint set of patterns in~$P_C$. This algorithm has linear running time in the number of intervals. It hence runs in $O(n^2)$ time on $P_C$. For completeness we summarize the algorithm below.

Each pattern $p \in P_C$ is defined by an interval $[i_p, \ldots j_p]$. We say that a pattern $p'$ is \emph{before} $p$, denoted by $p'\prec p$, if $j_{p'} < i_p$. The algorithm effectively follows a dynamic programming scheme, computing values $\Sigma_p$, expressing the MWIS including pattern $p$ on $\{ p \} \cup \{ p' \in P_C | p' \prec p \}$.
This can readily be expressed as $\Sigma_p = w_p + \max_{p' \prec p} \Sigma_{p'}$.
To compute the $\Sigma_p$ values efficiently, we sort the endpoints of all intervals (using Counting Sort in linear time), and sweep over this list.
During this sweep, we keep track of $\Sigma_*$, the MWIS on the intervals that have already ended; initially $\Sigma_* = 0$. At the start $i_p$ of pattern $p$, $\Sigma_p$ is hence computed as $w_p + \Sigma_*$ in constant time. At the end $j_p$, we update $\Sigma_*$ if $\Sigma_p$ exceeds it. As such, computing $\Sigma_p$ for all $p \in P_C$ takes linear time.
By keeping track of which pattern defines $\Sigma_*$, we store the choices made and recover the actual patterns of the MWIS in linear time. 

\mypar{Selecting bicliques and stars.} Once the cliques from $P_C$ are selected, we greedily try to add patterns from $P^*\setminus P_C$ to $P$: we iteratively add the largest biclique or star from $P_B \cup P_S$ that is disjoint from all patterns in $P$. Since $P_B \cup P_S$ contains $O(n^2)$ patterns, the straightforward implementation may require $O(n^4)$ pairwise disjointness checks. However, in practice, this worst-case behavior does not occur, as the number of candidates and the number of selected patterns in this category tends to be much smaller.

\mypar{Filtering small patterns.} 
In the selection procedure, we have the option to filter out candidate bicliques that adhere to our definition of noisy pattern, but that are still not deemed appropriate given the currently selected patterns. For example, for large graphs, we may want to filter out patterns that provide less than some small (constant) number of black-black adjacencies, that is, patterns with too low weight. Another option is to filter out patterns that have less weight than a chosen fraction of the highest weight pattern that is currently selected. For example, we could filter out all candidates with weight smaller than $1\%$ or $5\%$ of the highest weight pattern in~$P$. If we maintain the maximum weight of the patterns in~$P$, such filtering checks can easily be done when starting an iteration of he loop on line~\ref{alg:filter-line} of Algorithm~\ref{alg:pattern-select}.

\begin{algorithm}[t]
  \caption{\textsc{SelectDisjointPatterns}$(P^*)$}\label{alg:pattern-select}
  \begin{algorithmic}[1]
  \REQUIRE Candidate pattern set $P^* = P_C \cup P_B \cup P_S$
  \ENSURE Pairwise disjoint pattern set~$P$
  
  \smallskip 
\COMMENT{Select maximum weight disjoint cliques}
\STATE Let $\Gamma$ be the sorted endpoints of intervals in $P_C$
\STATE $\Sigma_* = 0$
\FOR{each endpoint $\iota$ in $\Gamma$}
    \IF{$\iota$ is left endpoint $i_p$}
        \STATE $\Sigma_p = w_p + \Sigma_*$
    \ELSIF{$\iota$ is right endpoint $j_p$ and $\Sigma_p > \Sigma_*$}
        \STATE $\Sigma_* = \Sigma_p$, \quad $p_* = p$
    \ENDIF
\ENDFOR
\STATE Add pattern $p_*$ to $P$
\STATE $\Sigma_* = \Sigma_* - w_*$
\STATE Let $\Gamma_b$ be right endpoints sorted right-to-left
\FOR{each endpoint $j_p$ in $\Gamma_b$}
    \IF{$j_p < i_{p_*}$ and $\Sigma_j == \Sigma_*$}
        \STATE Add $p$ to $P$, \quad $\Sigma_* = \Sigma_* - w_p$, \quad $p_* = p$
    \ENDIF
\ENDFOR
\COMMENT{Greedily add bicliques and stars}
\STATE Sort $P_B \cup P_S$ in order of descending weight
\FOR{each pattern $p$ in $P_B\cup P_S$} \label{alg:filter-line}
    \FOR{each pattern $p'$ in $P$}
        \IF{$p$ intersects $p'$}
            \STATE Skip to next iteration of outer loop
        \ENDIF
    \ENDFOR
    \STATE Add $p$ to $P$
\ENDFOR
\RETURN $P$
\end{algorithmic}
\end{algorithm}

\section{Visualizing Noisy Patterns}\label{sec:visualizing}
To visualize the noisy patterns $P$ detected by our pipeline, we introduce \emph{Ring Motifs}: a novel motif simplification that explicitly encodes the noise in graph patterns. In this section we describe the visual design of the motif glyphs and a layout algorithm to automatically generate Ring Motif simplifications.

\subsection{Visual Design}\label{sec:glyphs}
The goal of a Ring Motif simplification is to visualize the high-level structure of a graph, through a visual summary of the noisy patterns. Our summary uses three types of visual elements: clique glyphs, biclique (and star) glyphs, and links. We first discuss the pattern glyphs, and then focus on links. Throughout, as stars are a specific form of biclique, we treat bicliques and stars analogously and only discuss bicliques.

For our pattern glyphs, we assume that each visualized pattern $p\in P$ is assigned a color~$c_p$. Furthermore, let $\square_p$ be the set of cells of $M_\rho$ covered by $p$, and let $E_p$ be the set of black cells in $\square_p$.
For all patterns, the area of the colored region of their glyphs corresponds to the edges in the pattern. Vertices of a pattern map to their boundary, which is essential to our concept of links.

\setlength{\intextsep}{3pt}
\setlength{\columnsep}{9pt}
\begin{wrapfigure}{r}{0.19\columnwidth}
        \includegraphics[width=0.19\columnwidth]{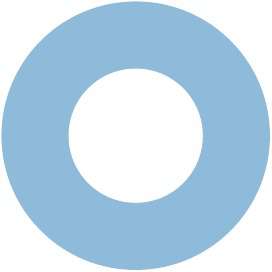}
\end{wrapfigure}
\mypar{Cliques} are visualized as (circular) annulus glyphs. For a clique pattern $p \in P$, the annulus glyph has color~$c_p$ and can be interpreted as a disk of area $|\square_p|$ proportional to the number of possible edges, with a hole of size $|\square_p|-|E_p|$, corresponding to the number of missing edges. The radius of the outer circle is therefore defined as $\sqrt{|\square_p| / \pi}$, and the radius of the inner circle is $\sqrt{(|\square_p| - |E_p|) / \pi}$. A noiseless pattern has a 0-radius hole and is therefore simply a solid disk.
To map the vertices of $p$ to the outer boundary, we partition the outer circle into equal arcs, and allocate an arc to each vertex in order of $\rho$.

\setlength{\intextsep}{3pt}
\setlength{\columnsep}{9pt}
\begin{wrapfigure}{r}{0.19\columnwidth}
    \centering
    \includegraphics[width=0.19\columnwidth]{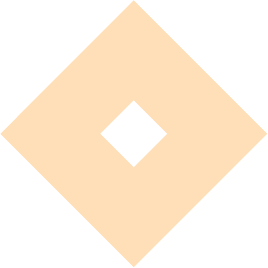}
\end{wrapfigure}
\mypar{Bicliques} are visualized as diamond annulus glyphs. For a biclique pattern $p\in P$, the diamond annulus has color~$c_p$ and can be interpreted as a diamond of area $|\square_p|$ with a hole of size $|\square_p|-|E_p|$. The side length of the outer square is defined as $\sqrt{|\square_p|}$, while the hole has side length $\sqrt{(|\square_p| - |E_p|)}$. A noiseless pattern is a solid diamond. Stars, being a special type of biclique, are visualized analogously.
To map the vertices of $p$ to the outer boundary, we assign the two sets to opposing sides of the diamond, partition them into equal segments and allocate a segment to each vertex in order of $\rho$.

\setlength{\intextsep}{3pt}
\setlength{\columnsep}{9pt}
\begin{wrapfigure}{r}{0.2\columnwidth}
    \centering
    \includegraphics[width=0.19\columnwidth]{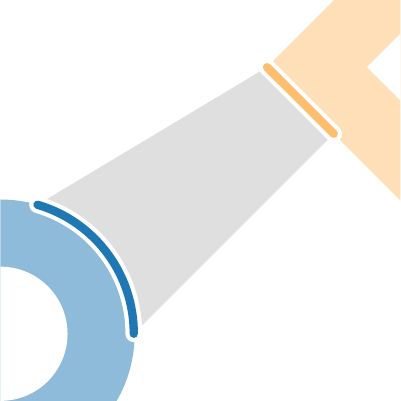}
\end{wrapfigure}
\mypar{Links} are gray trapezoids that show connections between patterns. Since we visualize a set~$P$ of pairwise disjoint patterns, none of our patterns share any edges. However, patterns in the same row or same column of matrix $M_\rho$ share vertices, and links visualize how vertices are shared between patterns: each link represents the vertex overlap between a pair of patterns, and hence connects along the boundary of the corresponding two glyphs, exactly where the shared vertices are mapped to their boundaries. Shared vertices are highlighted by a thicker outline where the link attaches.

To prevent clutter in the motif simplification, we do not place links between every pair of glyphs for patterns that share vertices. As our clique patterns are disjoint, they do not share vertices. However, bicliques may share vertices with cliques as well as other bicliques. Our links visualize only shared vertices between a  clique and a biclique, not between two bicliques. There are still visual cues that bicliques share vertices, if those vertices are also shared by a cliques: the attachment of their links then overlaps at the boundary of the clique. Links are slightly transparent to show more clearly when such overlaps happens.

\begin{figure*}
    \centering
    \includegraphics{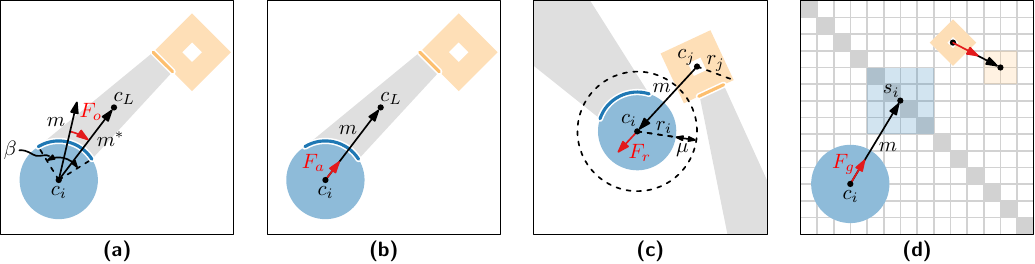}
    \caption{Ring Motifs force-based layout: \textbf{\textsf{(a)}} Rotational force, \textbf{\textsf{(b)}} Link attraction, \textbf{\textsf{(c)}} Glyph repulsion, and \textbf{\textsf{(d)}} Pattern gravity.}
    \label{fig:force-based}
\end{figure*}

By this choice, our Ring Motif simplification is also a direct generalization of the traditional node-link diagram: if every individual vertex is considered a ``clique'' and each edge a ``biclique'', then each vertex translates to a circular glyph and each edge to a diamond with two links connecting to its incident vertices. This is effectively a node-link diagram where each edge has one bend.


\setlength{\intextsep}{3pt}
\setlength{\columnsep}{9pt}
\begin{wrapfigure}{r}{0.06\columnwidth}
    \centering
    \includegraphics[]{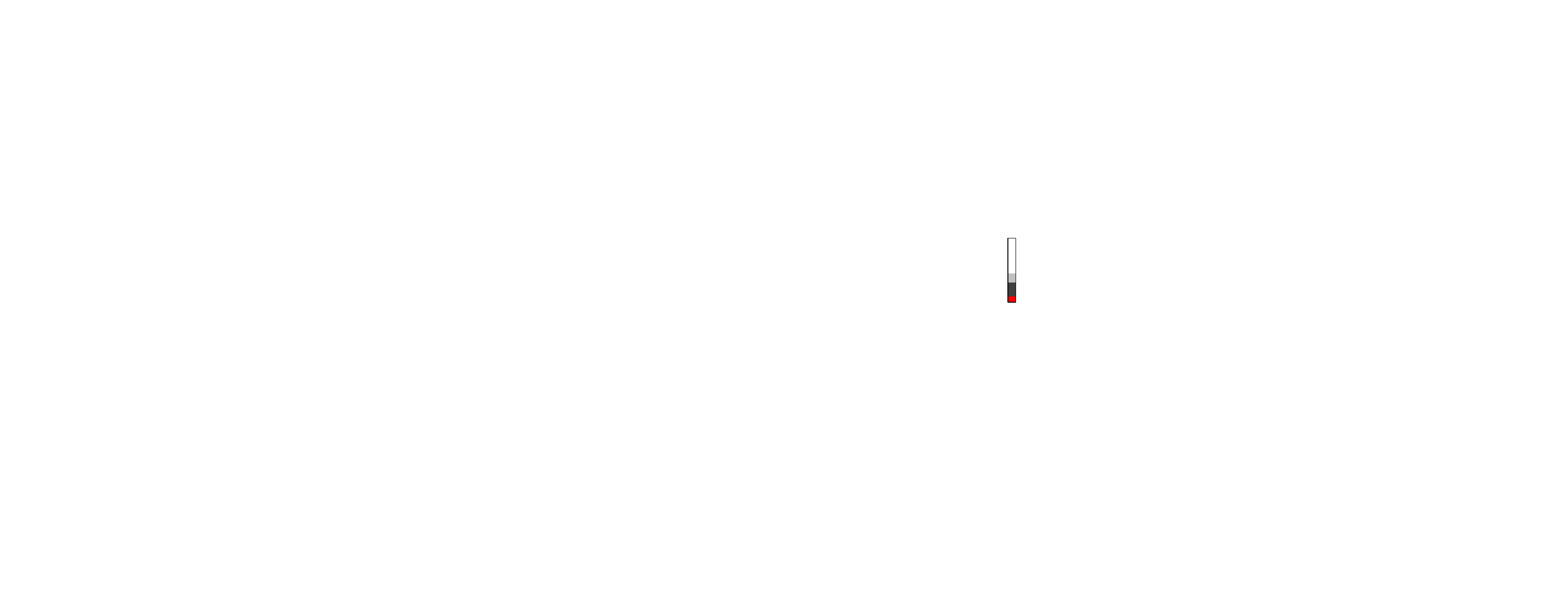}
\end{wrapfigure}
\mypar{Pattern precision bar.} We add an indicator of the precision of the visualized patterns~$P$ to our motif simplification, in the form of a stacked bar chart. This bar gives some insight into part of the data that cannot be derived from the motif simplification itself, namely what portion of the graph is actually visualized. The stacked bar chart consists of four bars, which together represent the cells of the adjacency matrix. The white and light gray bars together represent the white cells, while the dark gray and red bars represent the black cells. The white bar represents white cells that are not in any pattern and the light gray cells are white cells in a pattern. Similarly, the dark gray bar represents black cells in a pattern, and the red bar represents black cells that are not in any pattern. The size of each bar is proportional to the number of cells they represent. Importantly, the light gray bar represents the level of noise in~$P$ as a whole, while the red bar represents edges that are not represented by the motif simplification -- lower amounts of light gray and red hence are indicate a more accurate summary.

\subsection{Force-Based Layout}\label{sec:motif-alg}
To produce our Ring Motifs simplification we use a force-based layout. To initialize this iterative method, we place a glyph for each pattern $p$ at the centroid~$s$ of the coordinates of~$p$ (see Figure~\ref{fig:force-based}d), and attach corresponding links. Every iteration, we compute a rotation and a translation for each glyph; link positioning is directly derived from its attachments. The rotation and translation are computed through following four forces (see Figure~\ref{fig:force-based}).

\mypar{Rotational force.} To ensure that links attach to glyphs as straight on as possible, each link applies a rotational force on the glyphs it is attached to. Consider the glyph~$G_i$ for pattern~$p_i$ which is currently centered at position $c_i$ and rotated by angle $\alpha_i$. Furthermore, let there be a link~$L$ with centroid $c_L$ that attaches to this glyph at boundary $\theta(\alpha_i) = [a, b]$. Since $\theta$ has a fixed position along the boundary of~$G_i$, it is dependent on~$\alpha_i$.

To compute the rotational force $F_o(G_i, L)$, we consider three properties of $G_i$ and $L$. First, let $\beta$ be the central angle between $a$ and $b$, that is, the angle that $\theta$ spans around $c_i$. Second, let $m = (a+b)/2 - c_i$ be a vector from the center of~$G_i$ pointing to the middle of boundary $\theta$; if $\beta > \pi$ then we need to rotate $m$ by $\pi$ for it to point to the middle of $\theta$. Third, let $m^* = c_L - c_i$ be the vector pointing from the center of $G_i$ to the center of $L$. The rotational force is now defined as
\begin{align*}
    F_o(G_i, L) = c_o \cdot \angle(m^*, m) \cdot \frac{\beta}{2\pi}.
\end{align*}
Here, $c_o$ is a user parameter that allows for finetuning of the force, and the $\angle(m^*,m)$ function computes the smallest signed angle from vector $m$ to vector~$m^*$. The force~$F_o(G_i, C)$ hence pulls $m$ towards $m^*$ and is proportional to how much of the boundary of $G_i$ is covered by $\theta$: a larger boundary $\theta$ leads to a higher force.

\mypar{Link attraction.} Our links act similar to links in a force-based node-link layouts: the glyphs connected by a link are attracted to the link (and hence to each other). Again, consider a glyph $G_i$ with center~$c_i$ and a link $L$ attached to it, with centroid~$c_L$. Let $m = c_L - c_i$ be the vector pointing from the center of $G_i$ to the center of $L$, then the attraction force on $G_i$ is defined as
\begin{align*}
    F_a(G_i, L) = c_a \cdot \frac{m}{|m|}.
\end{align*}
We again introduce a user parameter~$c_a$ for tuning this force. Additionally, note that vector $m$ is normalized by dividing by the length $|m|$ of $m$, which ensures that the size of $L$ does not influence $F_a$.

\begin{figure*}[t]
    \centering
    \includegraphics[]{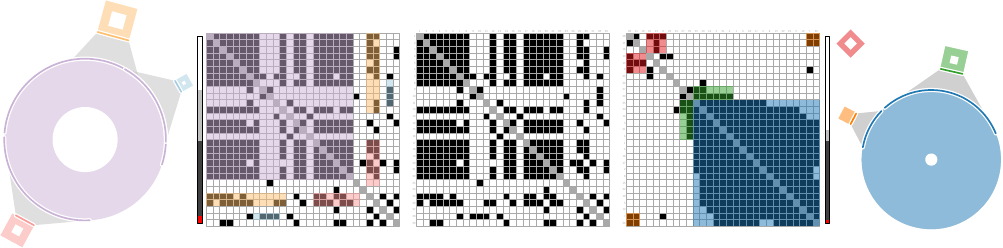}
    \caption{Matrix 1 of FLT data set. (Left) The matrix is not reordered and patterns are found with $\sigma = 0.2$ and $\tau=0.6$. (Right) The reordered matrix is optimal for Moran's I and patterns are found with $\sigma = 0.3$ and $\tau=1.0$. Patterns on the left miss more edges and are more noisy.}
    \label{fig:flt01}
\end{figure*}

\mypar{Glyph repulsion.} Between any two glyphs there is a repulsion force. This repulsion force is stronger when the glyphs are close and weaker when they are apart. Let $G_i$ and $G_j$ be two glyphs, with centers~$c_i$ and $c_j$ and (outer) radii~$r_i$ and $r_j$, respectively. Furthermore, let $m = c_i - c_j$ be the vector pointing from the center of $G_j$ to the center of $G_i$. The repulsion force $F_r(G_i, G_j)$ from $G_j$ onto $G_i$ is then defined as follows; $F_r(G_j, G_i)$ is defined analogously.
\begin{align*}
    F_r(G_i, G_j) = c_r \cdot \frac{m}{|m|} \cdot \left(\frac{r_i + r_j + \mu}{|m|}\right)^3
\end{align*}
This force has two user parameters~$c_r$ and $\mu$: $c_r$ acts as a tuning parameter for the repulsion force as a whole, but $\mu$ has a more subtle role. We normalize vector $m$ and scale it by a factor $((r_i+r_j+\mu)/|m|)^3$. This scaling factor is larger than one whenever the distance $|m|$ between the centers of $G_i$ and $G_j$ is smaller than $r_i+r_j+\mu$. If $\mu = 0$, this is when $G_i$ and $G_j$ overlap. The constant~$\mu$ therefore promotes a margin of size~$\mu$ between glyphs. Similarly, the scaling factor is smaller than one, and hence $F_r$ is weakened, when the distance between glyphs is larger than margin~$\mu$. By taking a (positive) power of the ratio between $r_i+r_j+\mu$ and $|m|$, in our case cubing the ratio, we strengthen the repulsion when glyphs are (close to) overlapping, and create a faster drop off of the force as glyphs move away from each other.

\mypar{Pattern gravity.} The final force ensures that the motif simplification remains compact, and that it has some coherence with the reordered matrix~$M_\rho$ that the visualized patterns are based on. In essence, every glyph~$G_i$ has a moderate pull towards its initial position, which was based on where pattern $p_i$ resides in the matrix $M_\rho$. Let $s_i$ be the initial position of~$G_i$, and $c_i$ its current center point. Then, let $m = s_i - c_i$ be the vector pointing from $G_i$ to its initial position. The generic gravitational force is then defined as
\begin{align*}
    F_g(G_i) = c_g \cdot \frac{m}{|m|}.
\end{align*}
Also for this force do we introduce a user parameter~$c_g$ for tuning. However, the force $F_g(G_i)$ is applied, as is, only to glyphs without attached links. For glyph with attached links, the gravity can interfere with the other forces, by pulling connected glyphs in different directions. To prevent this we do the following. For clique glyphs, we consider all clique glyphs that are reachable via links and other glyphs. For the clique glyphs that are in such a connected component, we compute the average starting position $s^*$ and the average~$c^*$ of their current center points. We then find the vector $m^* = s^* - c^*$ and use the normalized $m^*$ instead of $m$ for all the clique glyphs in the component.
For bicliques with attached links, we simply scale down the force by using $F_g(G_i) / 5$.

\mypar{Combining the forces.} For each glyph $G_i$, we compute the sum $\alpha^*_i = \Sigma_L F_o(G_i,L)$ of all the rotational forces of the links $L$ that attach to $G_i$. 
We update the rotation $G_i$ by adding $\alpha^*_i / r_i$. Note that we divide by the radius~$r_i$ of $G_i$, such that larger glyphs are less effected by the forces. Similarly, we compute the sum $\Sigma_L F_a(G_i, L) + \Sigma_{G_j} F_r(G_i,G_j)$ of all attraction and repulsion forces on $G_i$ by attached links. We add this sum to $F_g(G_i)$ to find the total translation~$c^*_i$ of $G_i$. Again we divide this translation by the radius of $G_i$ and add $c^*_i / r_i$ to $c_i$ to find $G_i$'s new position.

\section{Case Study}\label{sec:case-study}

We demonstrate and evaluate our noisy pattern pipeline and our Ring Motifs simplification. We use four real-life data sets that arise from different domains and exhibit varying structural properties. 



\begin{description}
    \item[FLT:] The ``flashtap'' data set used for MultiPiles~\cite{multipiles}. It is a temporal data set consisting of 96 time steps, each representing functional brain connectivity measured in a Parkinson’s disease study. The graphs each have 29 vertices and are rather dense.
    \item[ZKC:] A social network of karate club members, studied by Zachary~\cite{zachary-karate}. The data set has 34 vertices, representing the members of the club, and the edges model social interactions of club member outside of their club meetings.
    \item[SCH:] Temporal network data representing social interaction between children and teachers at a primary school~\cite{Gemmetto2014, 10.1371/journal.pone.0023176}. The data set is split into 1-hour intervals, resulting in 17 graphs. Each graph has 242 vertices and is fairly sparse.
\end{description}

\mypar{Implementation.}
A prototype implementation of the pipeline and force-based layout written in Java 21 is available online~\cite{code}. The matrix reordering step of the pipeline utilizes the Concorde TSP solver~\cite{concorde} on the NEOS server~\cite{neos} via XML-RPC. All figures in Section~\ref{sec:case-study}, as well as Figure~\ref{fig:teaser} are produced using this implementation; in some cases with slight postprocessing (such as translations/rotations). The pipeline parameter settings are denoted for each figure, while the force-based layout is generally ran with default parameters ($c_o = 0.8$, $c_a = 1, c_r=1, c_g = 1, \mu = 3$) unless the results are unsatisfactory --- if necessary, we slightly increased $c_r$ to prevent overlaps or $c_g$ for convergence, up to at most~2.

\begin{figure*}[t]
    \centering
    \includegraphics[width=\linewidth]{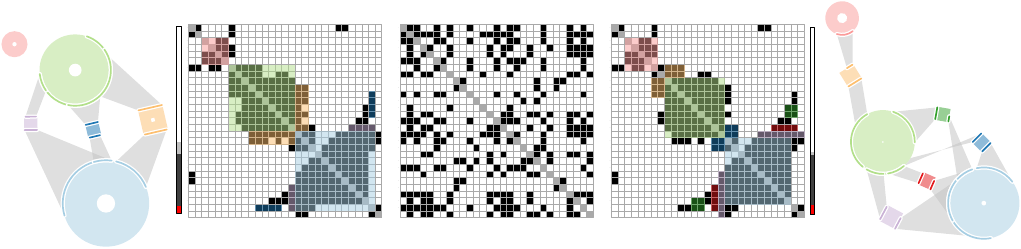}
    \caption{Matrix 58 of the FLT data set. (Left) Patterns are found with $\sigma = 0.5$ and $\tau=0.85$. (Right) Patterns are found with $\sigma = 0.6$ and $\tau=0.95$. Higher parameter settings lead to less noise in the patterns, more missed edges, and the Ring Motifs become more complex.}
    \label{fig:flt58}
\end{figure*}

\begin{figure}[b]
    \centering
    \includegraphics[width=\linewidth]{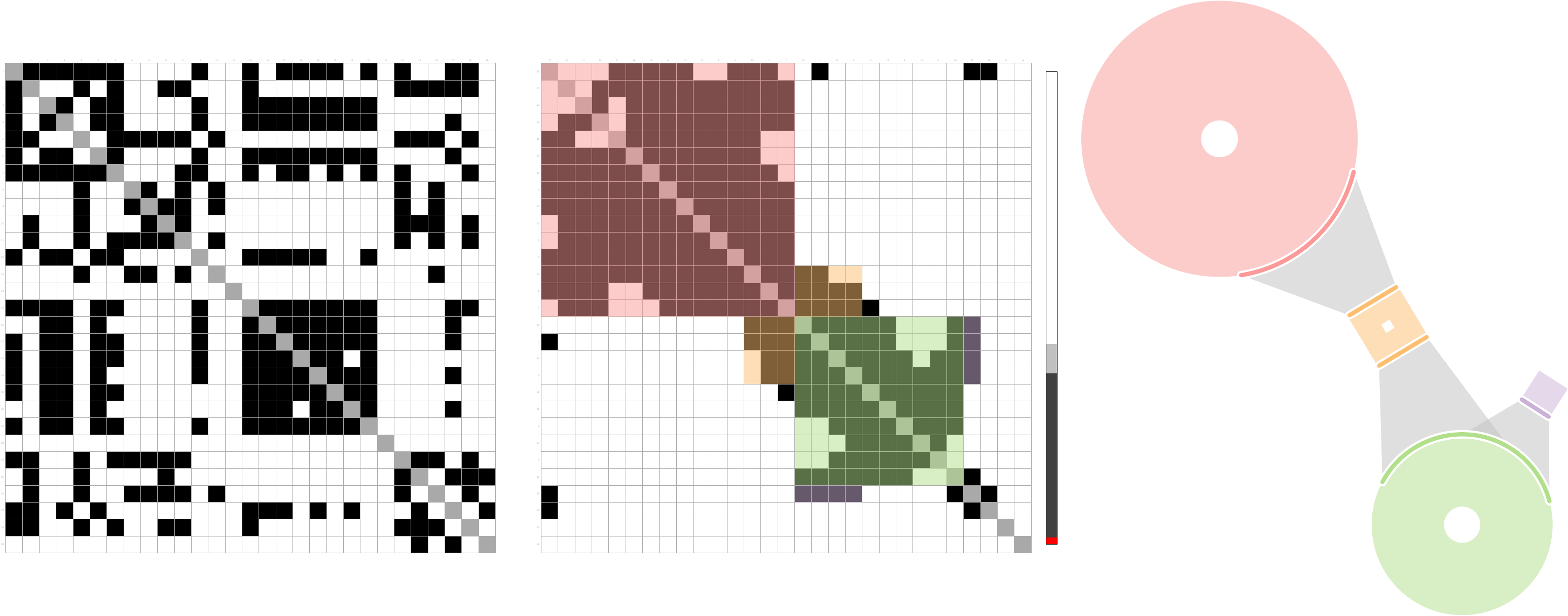}
    \caption{Matrix 35 of FLT data set with $\sigma = 0.5$ and $\tau=0.9$.}
    \label{fig:flt35}
\end{figure}

\subsection{Results}

\mypar{Running times.} Computationally, the matrix reordering step of the pipeline is the most demanding. Even so, the NEOS server finds a TSP tour within 20s, even on the SCH instances. All other parts of the pipeline are executed on a modern laptop (Intel Core i7-12700H, 2300Mhz, 14 Cores, running Windows 11) and finish almost instantaneously: the pattern enumeration and selection steps usually take <1s to finish. The force-directed layout is extremely light-weight; for our data sets, over 1000 iterations can be computed in a second. The layout usually converges within seconds.

\mypar{FLT.} The FLT data set consists of many graphs, and we use several of these graphs to show the effect the different parts of our pipeline have on the outcome of the pattern detection.

First, consider Figure~\ref{fig:flt01}, in which we show two runs of our pipeline on the first matrix in the FLT data set; one without reordering the matrix to optimize for Moran's $I$ (left) and one run where we use the full pipeline, including matrix reordering (right). Clearly, the patterns on the left are more noisy, as witnessed by the numerous white matrix cells that are covered and the larger holes in the Ring Motifs glyphs. This sentiment is echoed by the pattern precision bar: the left bar shows a larger light gray bar (more noise) and a larger red bar (less edges covered). This reinforces that optimizing for Moran's $I$ is crucial for pattern detection. 

Next, we visualize the 58th matrix of the FLT data set in Figure~\ref{fig:flt58}. Here we see a clear trade-off between the level of noise the patterns allow, and the coverage of the patterns. When the parameter settings are more strict, that is $\sigma$ and $\tau$ are closer to one, then the patterns contain less noise, but miss more edges. We also see that the patterns become smaller and the number of patterns increases. This also affects the Ring Motifs simplification: it must now summarize a more complex pattern decomposition. In Appendix~\ref{app:parameter}, we outline the effects of the parameters on the detected patterns for this data set; we provide a full grid of $\sigma$ and $\tau$ combinations to showcase patterns for varying parameters.

Finally, consider Figure~\ref{fig:flt35}, visualizing the 35th matrix in the FLT data set. While the initial matrix already shows plenty of rectangular black patterns, after reordering almost all edges merge into a single black shape. While this shape is not obviously rectangular in form, our pipeline can identify few noisy patterns to explain essentially the entire graph. Notice how noisy patterns are crucial in detecting this high-level graph structure.

\begin{figure}[b]
    \centering
    \includegraphics[width=\linewidth]{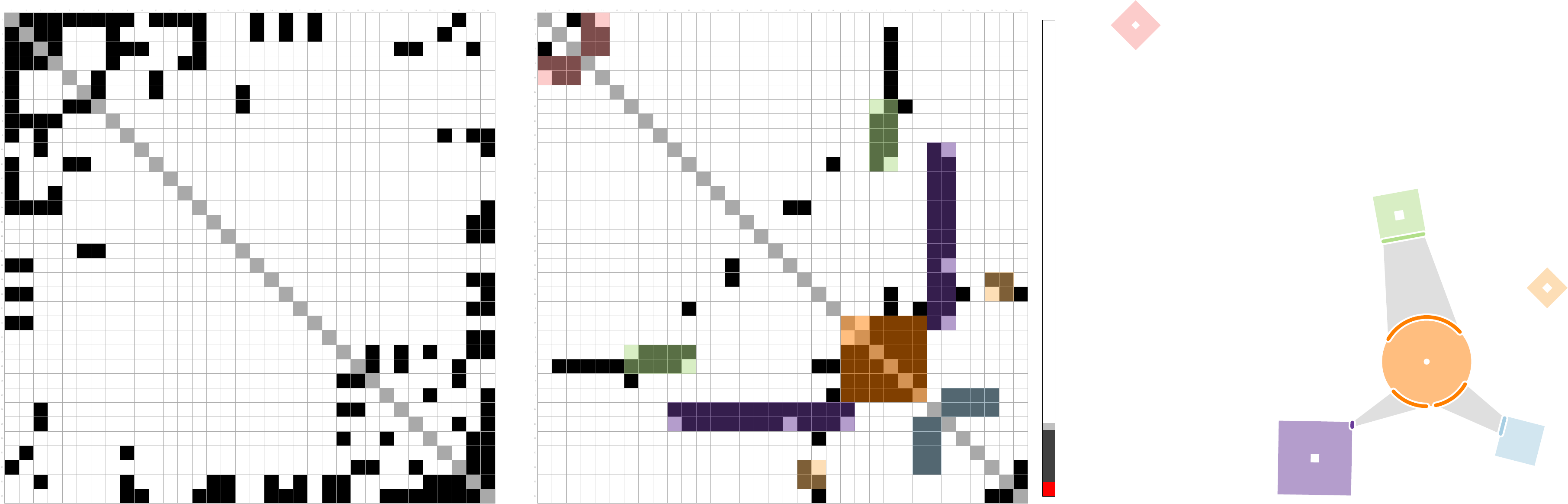}
    \caption{The ZKC data sets with parameters $\sigma = 0.5$ and $\tau=0.95$.}
    \label{fig:zkc}
\end{figure}

\mypar{ZKC.}
The Zachary's karate club data set has more predictable structures than the FLT data set. We use it to show that our pipeline can detect the known structures in this data set; see Figure~\ref{fig:zkc}.\newpage

\begin{figure*}[t]
    \centering
    \includegraphics[width=0.75\linewidth]{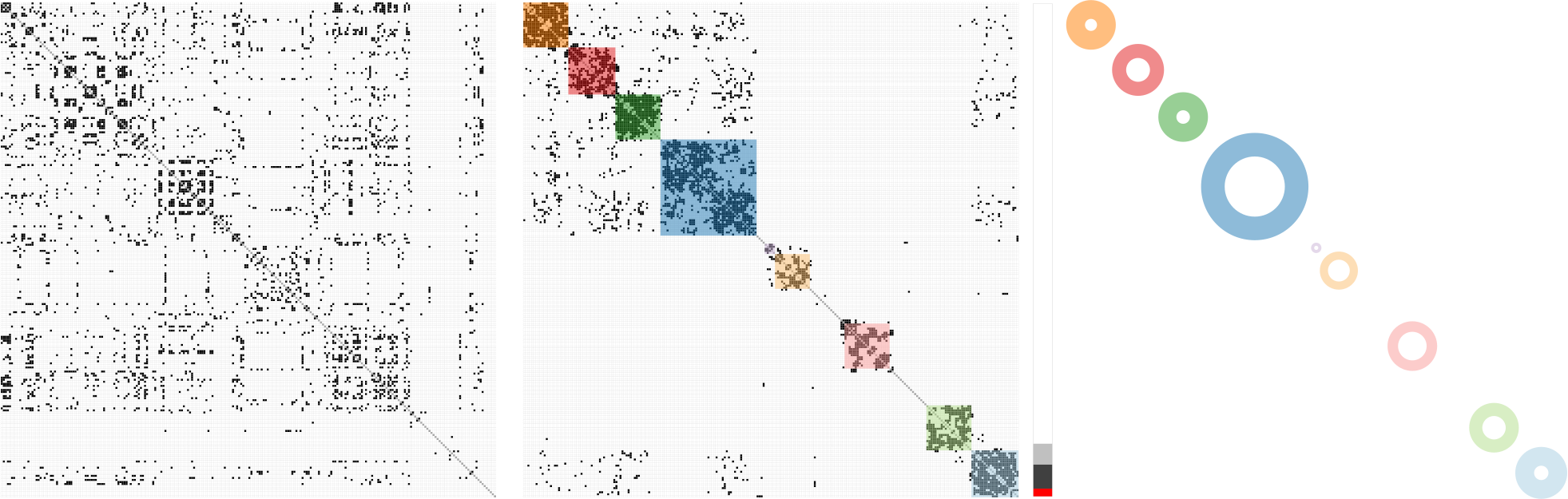}
    
    \vspace{10pt}
    \includegraphics[width=0.75\linewidth]{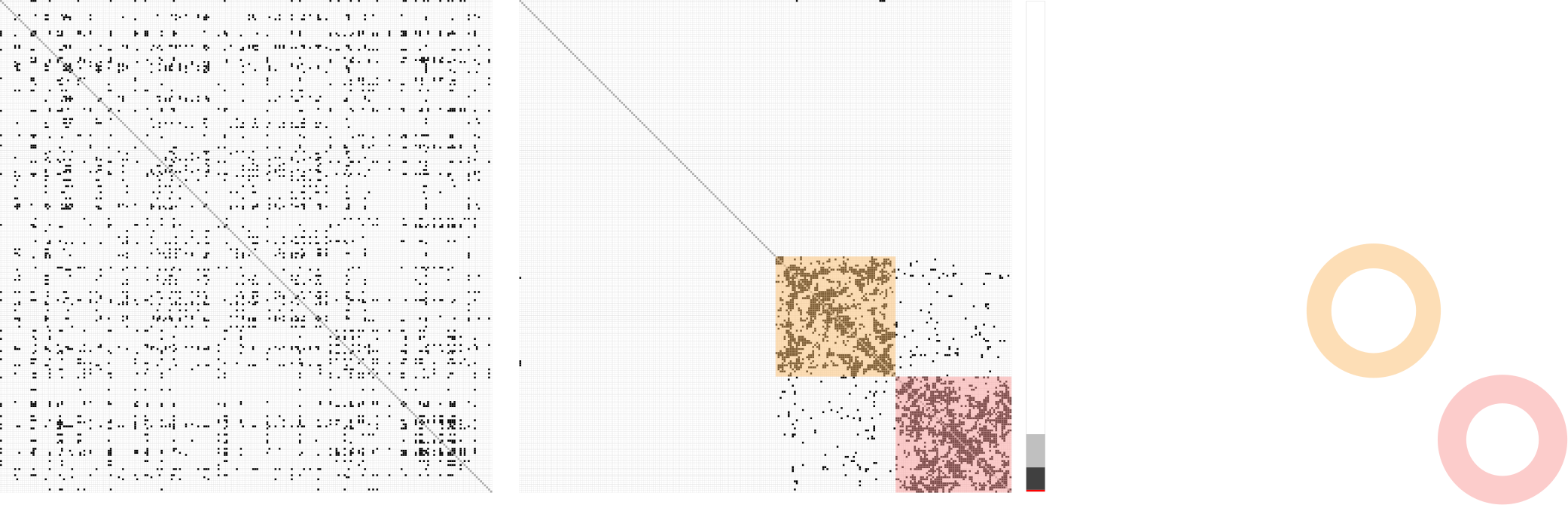}
    \caption{SCH graphs 8 (top) and 14 (bottom) with pattern parameters $\sigma = 0.2$ and $\tau=0.93$ and $\sigma = 0.1$ and $\tau=0.95$, respectively.}
    \label{fig:sch08+sch14}
\end{figure*}

In the ZKC data set, vertices 1 and 34 represent prominent members of the club, namely the karate instructor and the club president, respectively. It is known that a conflict between these members resulted in a split of the club, dividing members into two factions. The patterns found by our pipeline and the corresponding Ring Motifs capture many of the social interactions within the factions: The purple biclique captures most of the interactions within the club president's faction, as vertices 34 (the president) and 33 (likely his associate) have interactions with many people in their faction. The vertices that form the other side of the biclique have none, or very few interactions with any other club members. The orange clique and the green and red bicliques cover most of the interactions within the faction of vertex 1 (the instructor). 
Though the factions are disjoint, there is an outlying link between them, which, due to the noise parameters, can be captured in a pattern, visually linking the factions in the Ring Motif simplification.

\mypar{SCH.} The data set on social interaction in a primary school has more vertices and less clear-cut structures than the other data sets. Here, we want to showcase how our pipeline can reveal noisy graph patterns that are not directly apparent. Since the graph is large, we filter out small patterns with less than 1\% of the largest selected pattern weight. Figure~\ref{fig:sch08+sch14} visualizes the 8th and 14th SCH graph. 

For the 8th graph, there seems to be little more structure than some small bicliques, when considering the input matrix. However, our pipeline finds several reasonably sized noisy cliques. The reordering also showed that many vertices have virtually no edges, or only within their clique. For the 14th graph in the data set, the initial graph is even more unstructured than for the 8th graph. It is therefore extra surprising to see that most vertices have no incident edges. The existing edges cluster together into two noisy cliques that our pipeline accurately identifies. One thing to note here, is that for sparse matrices, the $\tau$ parameter should be kept high, to ensure tight patterns. By decreasing the $\sigma$ parameter can we still allow for increased amounts of noise to find the patterns in this data set.

\section{Conclusion}\label{sec:conclusion}
In this paper, we introduced noisy graph patterns as a way of summarizing the high-level structures in a graph. We developed a pipeline for detecting these noisy patterns, which emerge in well-ordered matrices as contiguous rectangles with sufficiently many black cells and black-black adjacencies. Hence, our pipeline first reorders the adjacency matrix of the input graph, to optimize for Moran's $I$ and create blocks of black cells. We then find candidate patterns, cliques and bicliques, as square submatrices along the diagonal and off-diagonal (rectangular) submatrices, respectively. Out of these candidates, the pipeline selects a maximal set of disjoint patterns; adding other candidates creates overlap. To the best of our knowledge, this way of defining and detecting noisy patterns via Moran's~$I$ is unique to our approach.

To visualize the patterns found by our pipeline, we proposed the Ring Motif simplification. In this motif simplification, clique and biclique patterns are visualized using glyphs that explicitly encode noise: the glyphs have a hole proportional to the noise in the pattern, resulting in ring-like shapes. Using several real-world data sets, we show that the combination of our pipeline and Ring Motifs is able to reliably detect and visualize high-level graph structures.

There are plenty of avenues for future research. The most obvious direction, is looking for more applications of our pipeline. The detection of noisy patterns can be a versatile preprocessing step for other graph analysis and visualization techniques. For example, the motif simplification used in BioFabrics~\cite{DBLP:conf/vis/FuchsD25} currently lacks algorithms to automatically find patterns. Additionally, graph visualization techniques that require some predefined grouping on the vertices or edges of the graph~\cite{DBLP:journals/tvcg/JianuRHT14,DBLP:journals/cgf/VehlowBW17}, can be automated for graphs without grouping, by linking up with our pipeline.

Finally, several parts of our pipeline and Ring Motifs can be further refined. For example, the enumeration and selection procedures could be revised to improve the resulting noisy-pattern decomposition. Additionally, the NEOS server has a limit of 16MB on the input size, making it unable to deal with large (1600+ vertex) graphs; other TSP solutions, such as approximations or heuristics, must be investigated for large graphs. For the Ring Motifs, the force-based layout can be refined, for example by rotating link attachments separately. This makes the layout less rigid and probably increases coherence with the matrix ordering. Furthermore, the pattern gravity mainly serves as a way to ensure coherence with the matrix view. Coherence may be less of a concern when Ring Motifs are used as a stand-alone visualization. The connected components can then easily be laid out differently, for example, ordered by their size. Lastly, the usability of Ring Motifs should be further studied via user experiments.

\mypar{Acknowledgments.} The authors would like to thank Nathan van Beusekom and Rein Buseman for initial explorations on the topic of this paper.

\bibliographystyle{eg-alpha-doi} 
\bibliography{references}       


\newpage
\appendix
\section{Parameter Experiment}
\label{app:parameter}
In the main paper, we propose a pipeline to detect noisy graph patterns by optimizing the ordering of a matrix for Moran's~$I$ and then selecting rectangular submatrices as patterns. We introduce two parameters, $\sigma$ and $\tau$ that allow the user to tune the level of noise in the detected patterns. Here we give a more comprehensive overview of the effect these parameters have on the selected patters: in Table~\ref{tab:parameter-experiment} we show the detected patterns for different combinations of $\sigma$ and $\tau$. Both parameters take values between 0 and 1. Low values of both $\sigma$ and $\tau$ lead to very noisy patterns. As either parameter increases towards 1, the patterns become less noisy, and as a result also smaller, in case the matrix is sparse, as is the case here. When both $\sigma$ and $\tau$ approach 1, the detected patterns become pure patterns, that is, fully black rectangular submatrices without any missing edges.

Somewhere in the middle between these two extremes, one finds the most interesting noisy pattern decompositions of the graph. In the middle of the grid, we clearly see patterns with relatively few white cells, that manage to leave very few black cells uncovered. These combinations of patterns hence give a good overview of the graph as a whole, while requiring only a moderate amount of noise. Though the ``optimal'' parameter choice is very data set and use case dependent, we suggest to default to high values of $\tau$, for example $\tau = 0.9$ and use $\sigma$ as a way of controlling the noise: for low noise choose a high value, such as $\sigma \geq 0.7$, and for larger noisy patterns, choose $\sigma\leq 0.5$.

\begin{table*}[]
    \centering
    \caption{Parameter experiment for matrix 58 in the FLT data set; $\sigma$ ranges from 0.1 to 0.9 and $\tau$ ranges from 0.5 to 1.0}
    \begin{tabular}{cccccc}
        & $\sigma = 0.1$ & $\sigma = 0.3$ & $\sigma = 0.5$ & $\sigma = 0.7$ & $\sigma = 0.9$ \\
        \raisebox{.08\linewidth}[0pt][0pt]{\rotatebox[origin=c]{90}{$\tau = 0.5$}} & \includegraphics[width=0.16\linewidth]{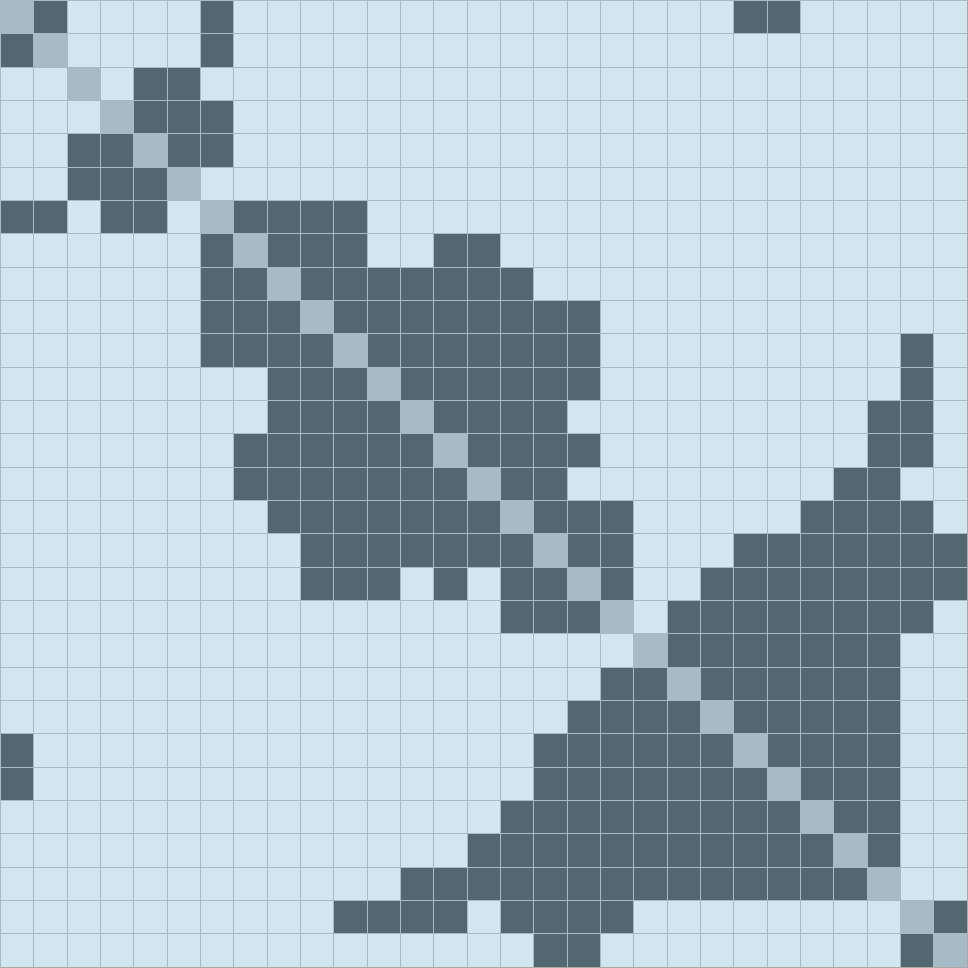} & \includegraphics[width=0.16\linewidth]{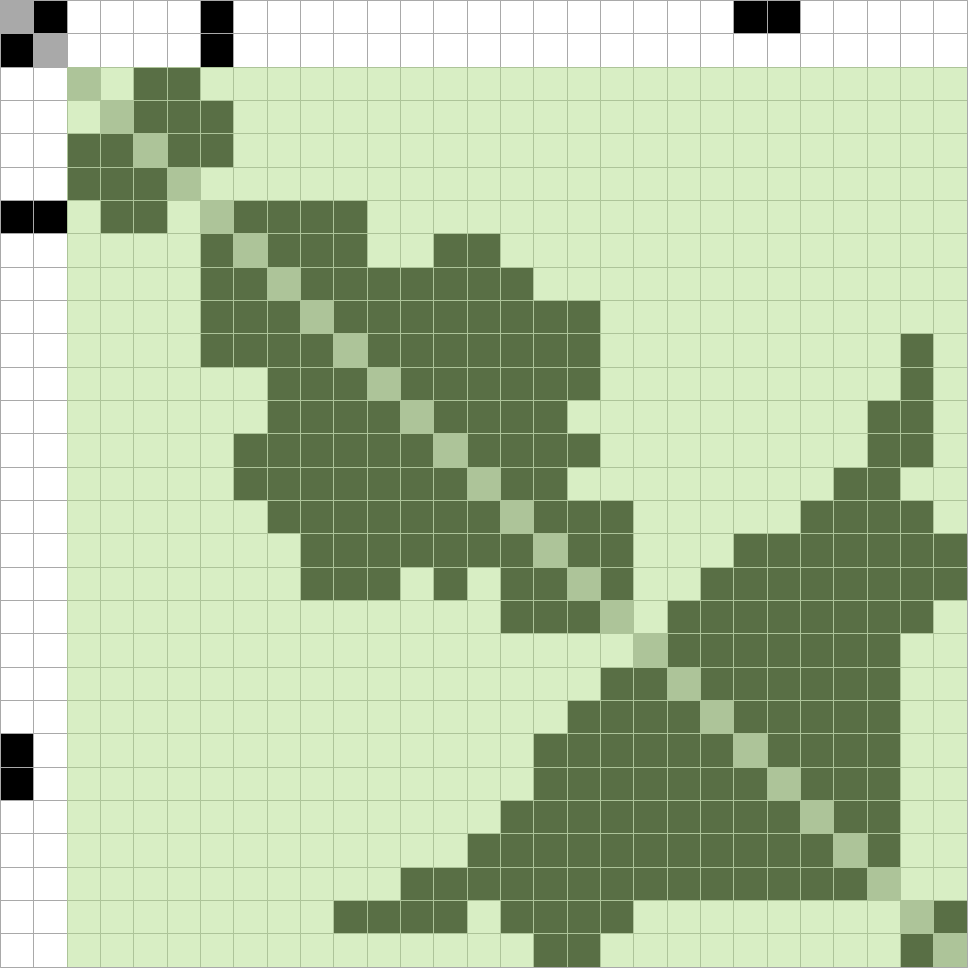} & \includegraphics[width=0.16\linewidth]{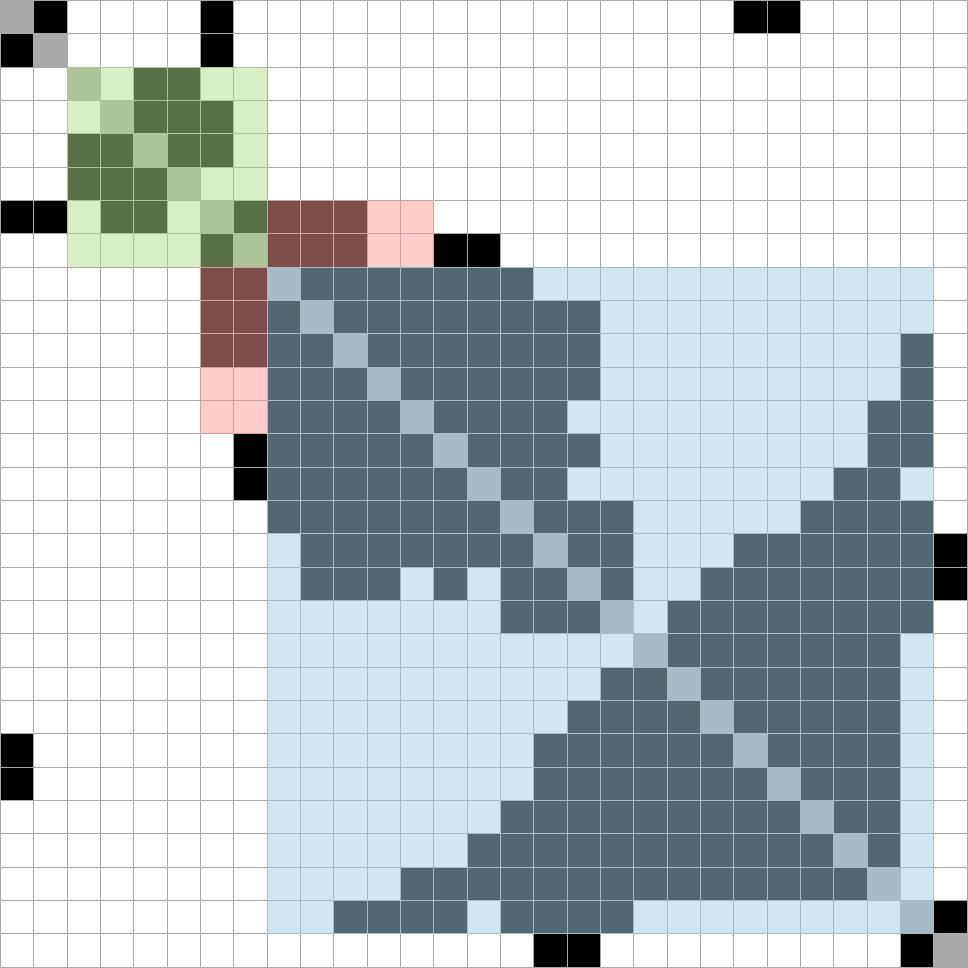} & \includegraphics[width=0.16\linewidth]{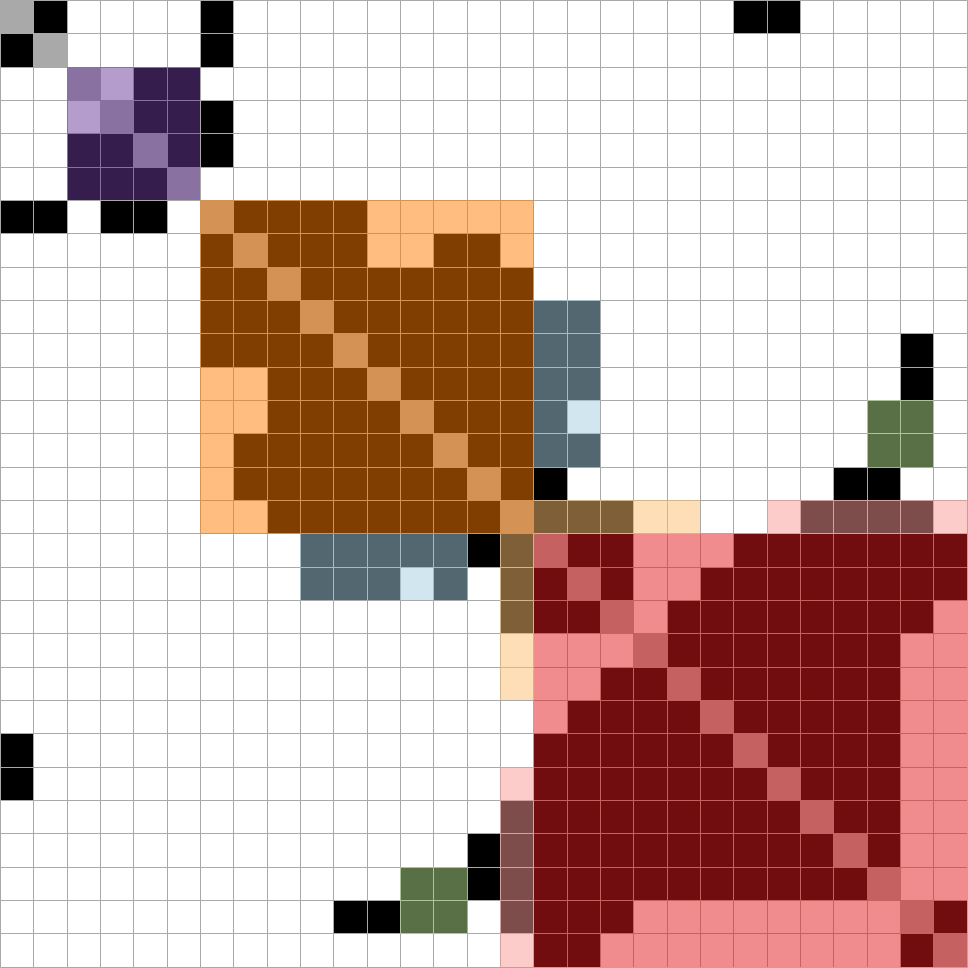} & \includegraphics[width=0.16\linewidth]{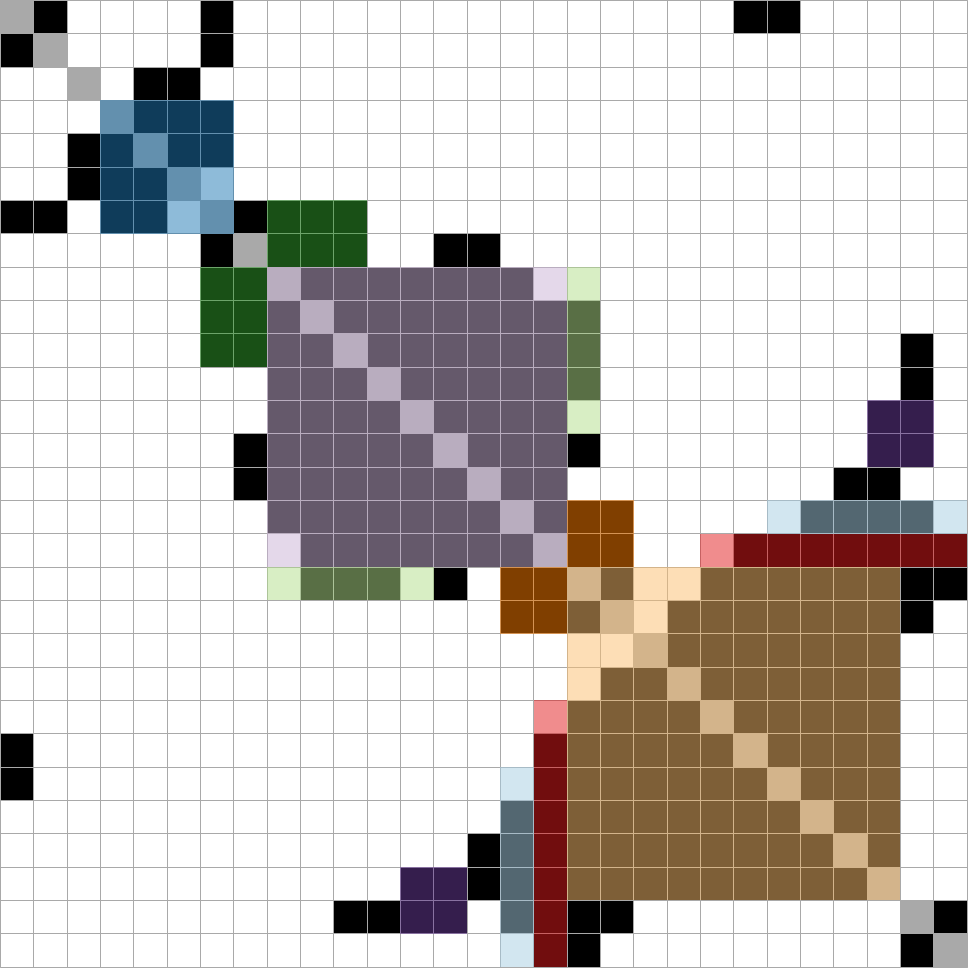} \\
        \raisebox{.08\linewidth}[0pt][0pt]{\rotatebox[origin=c]{90}{$\tau = 0.6$}} & \includegraphics[width=0.16\linewidth]{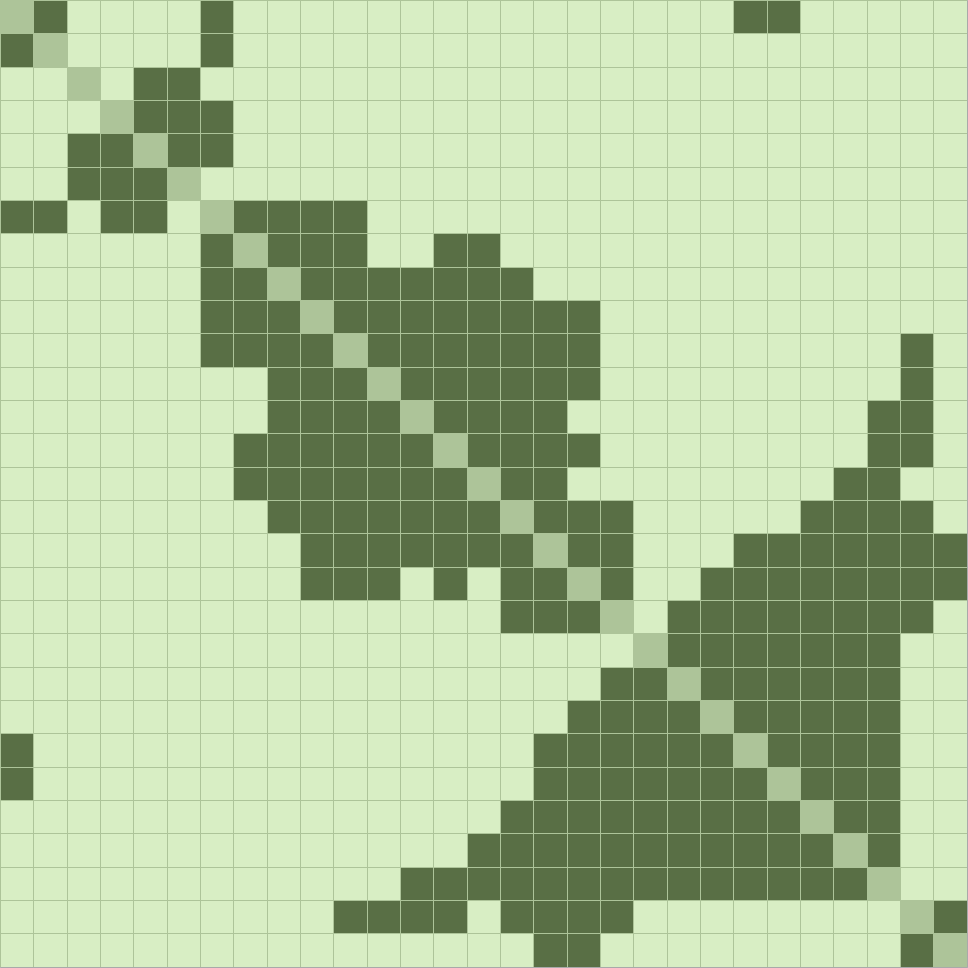} & \includegraphics[width=0.16\linewidth]{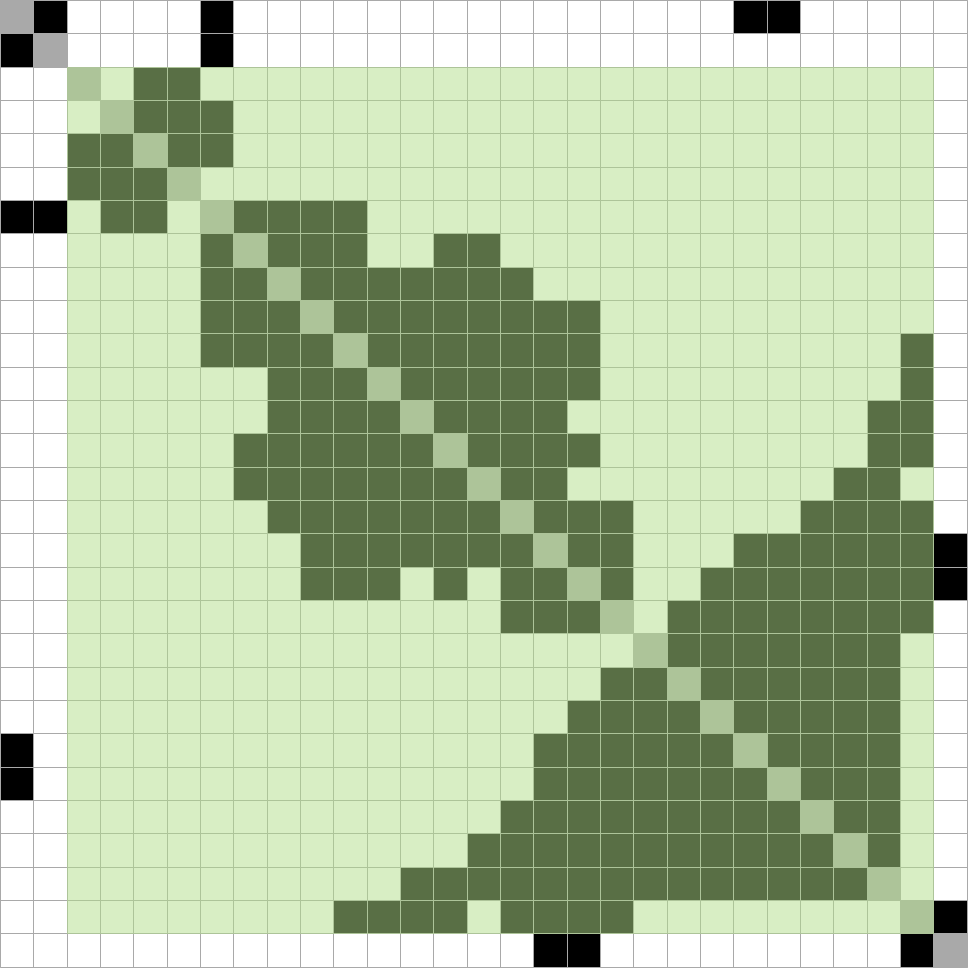} & \includegraphics[width=0.16\linewidth]{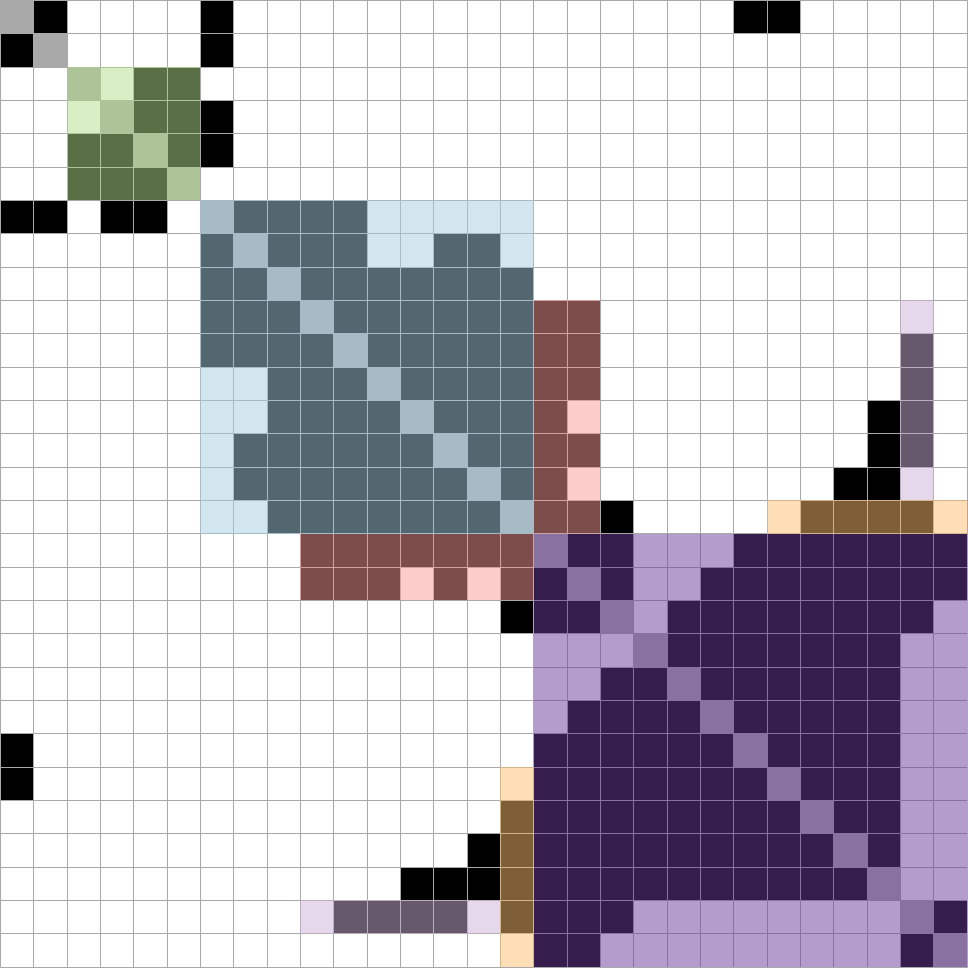} & \includegraphics[width=0.16\linewidth]{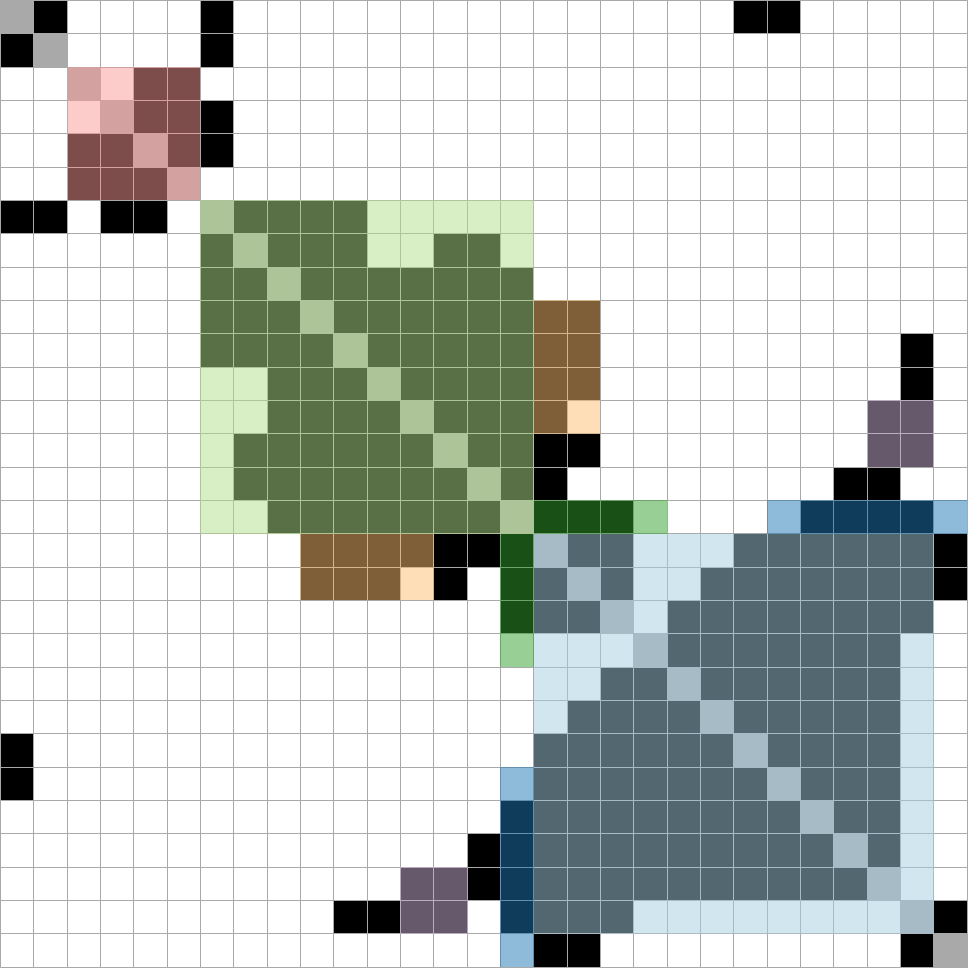} & \includegraphics[width=0.16\linewidth]{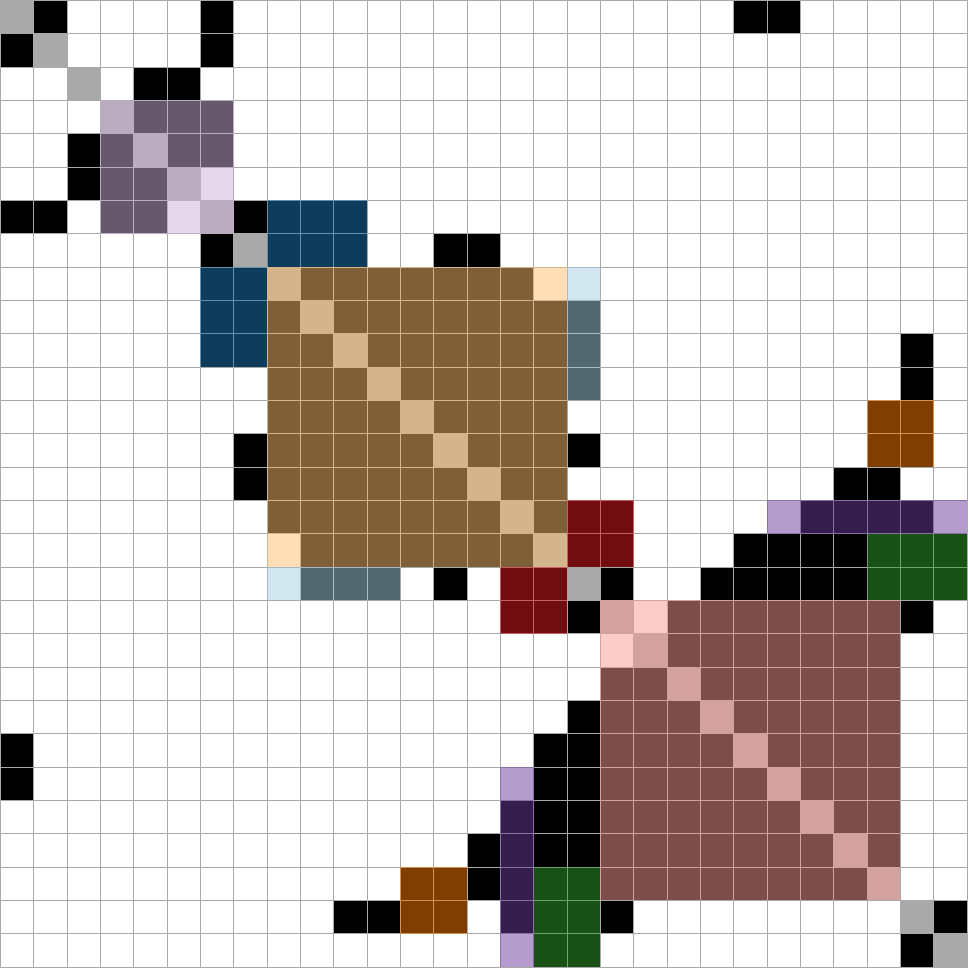} \\
        \raisebox{.08\linewidth}[0pt][0pt]{\rotatebox[origin=c]{90}{$\tau = 0.7$}} & \includegraphics[width=0.16\linewidth]{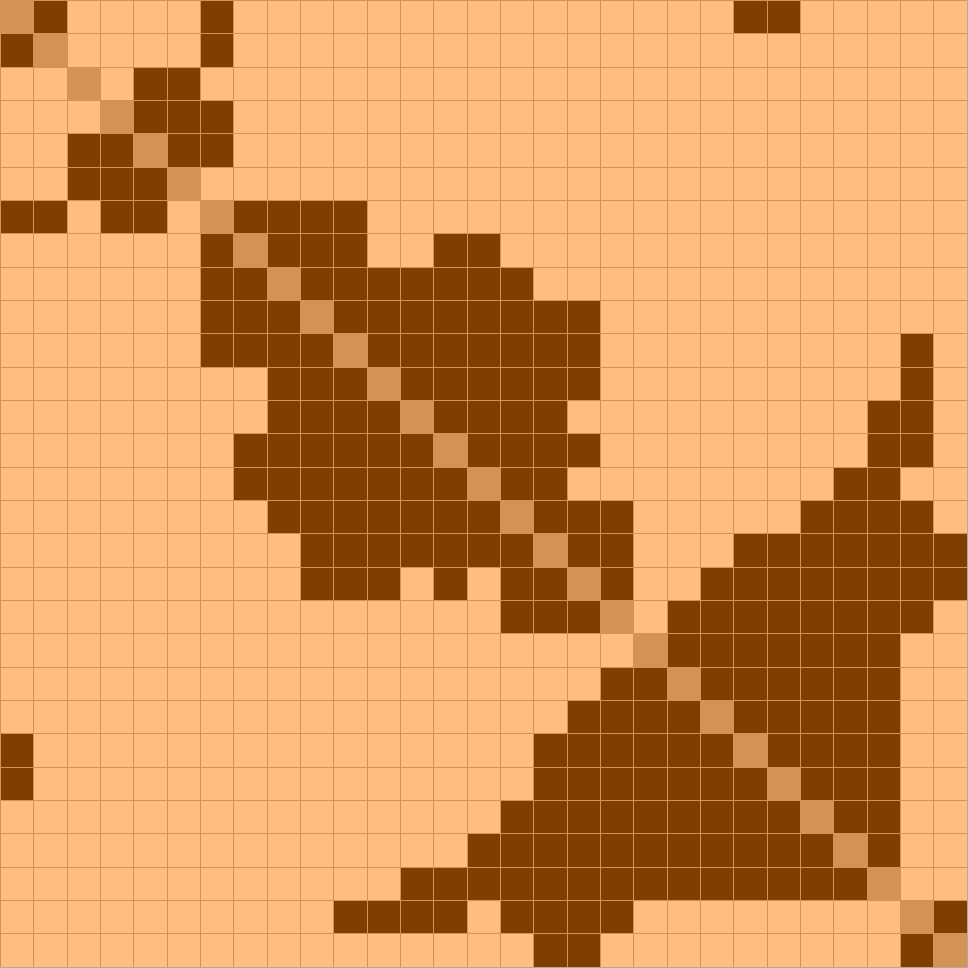} & \includegraphics[width=0.16\linewidth]{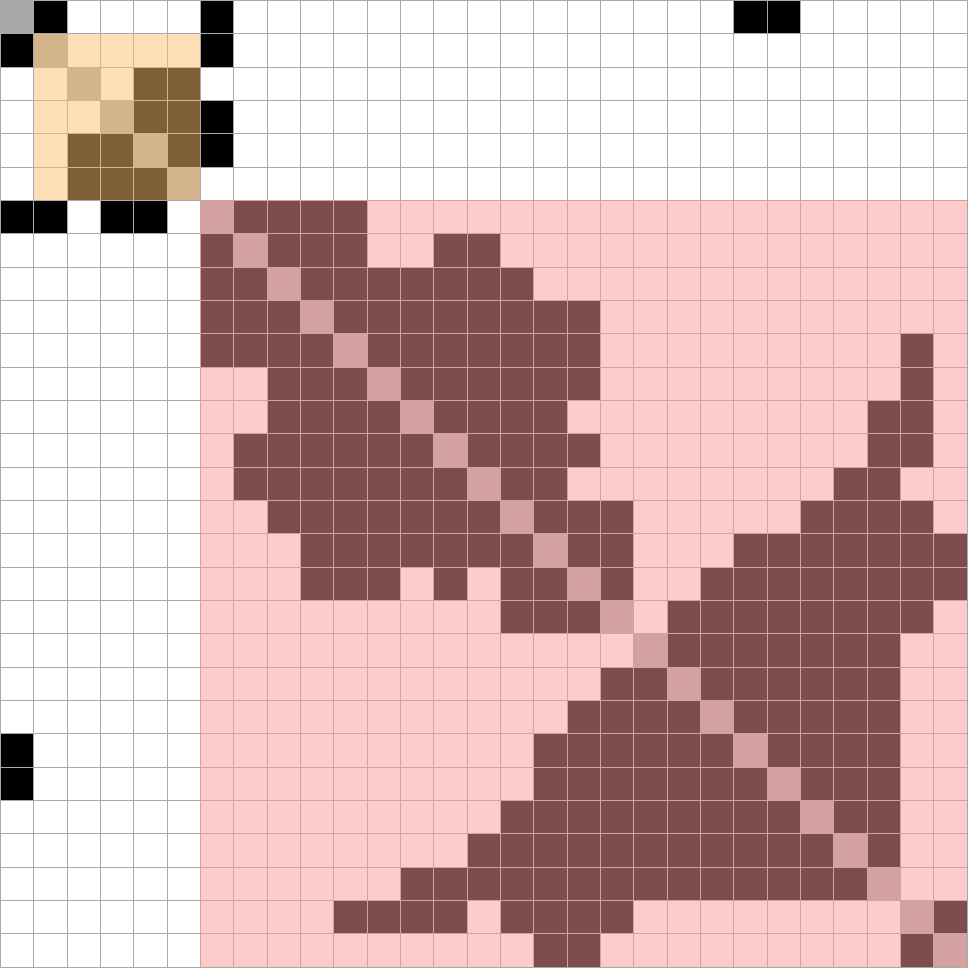} & \includegraphics[width=0.16\linewidth]{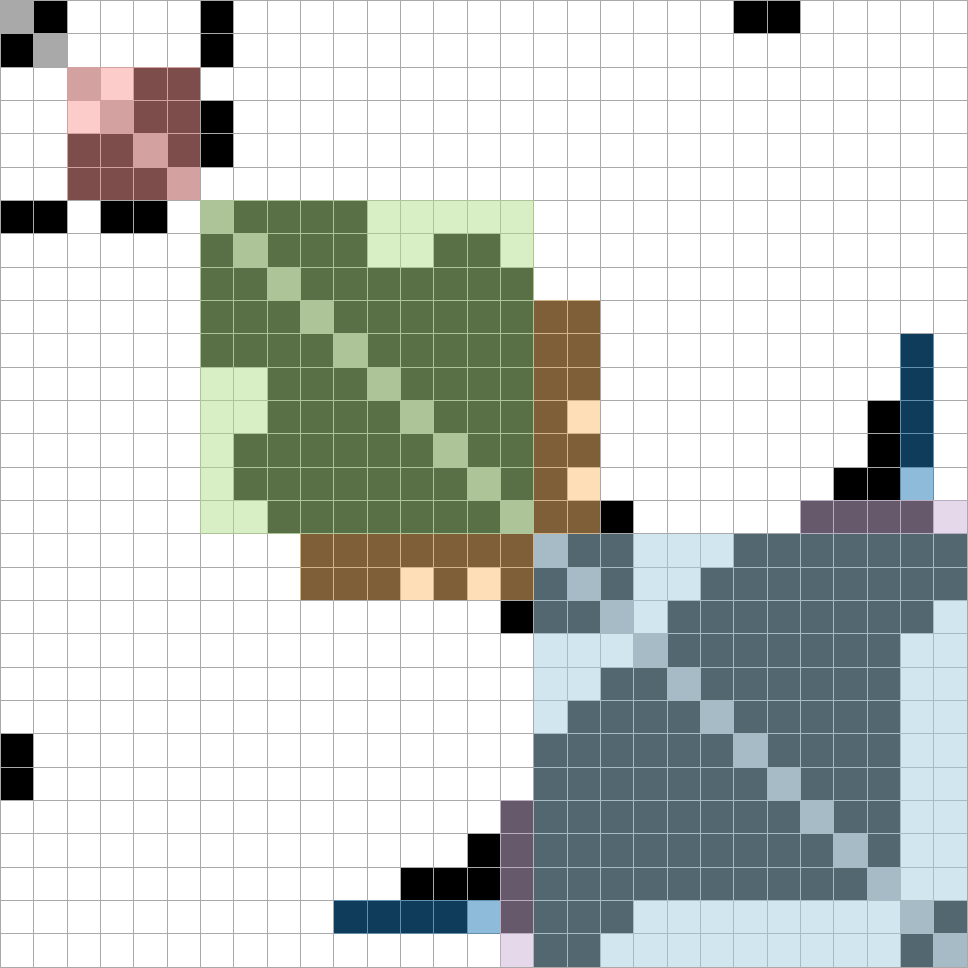} & \includegraphics[width=0.16\linewidth]{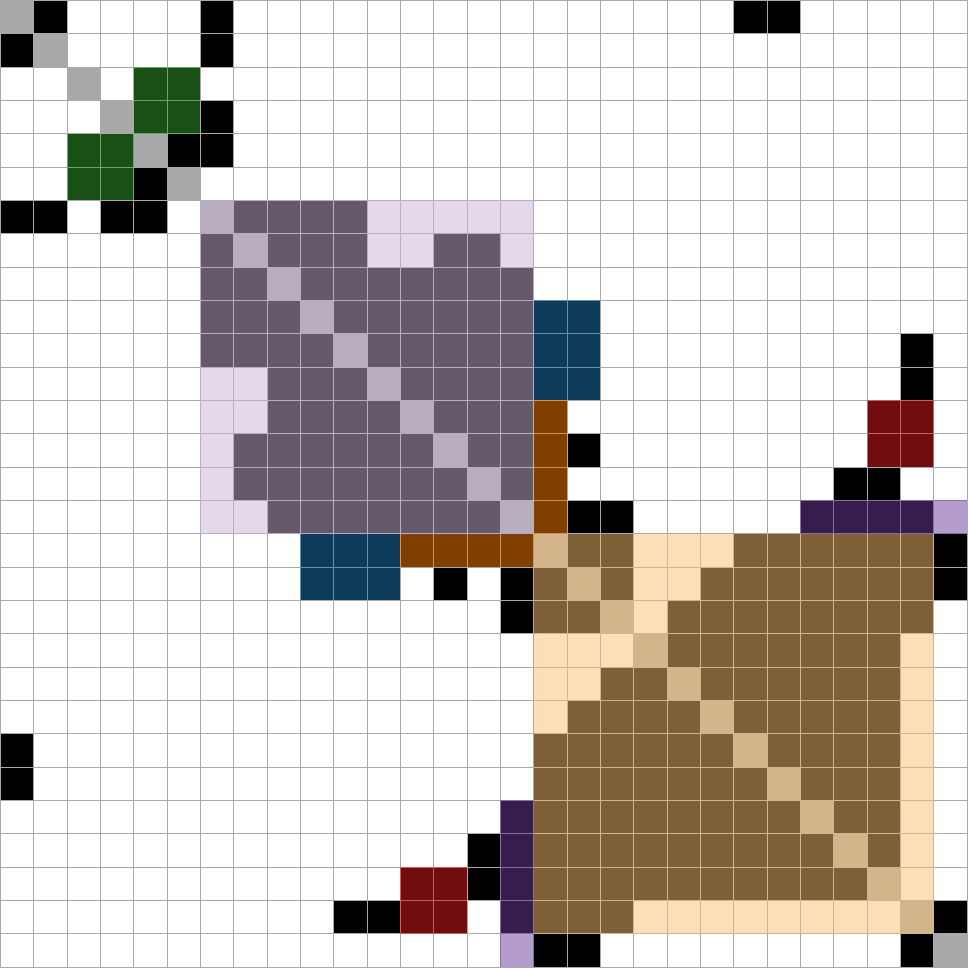} & \includegraphics[width=0.16\linewidth]{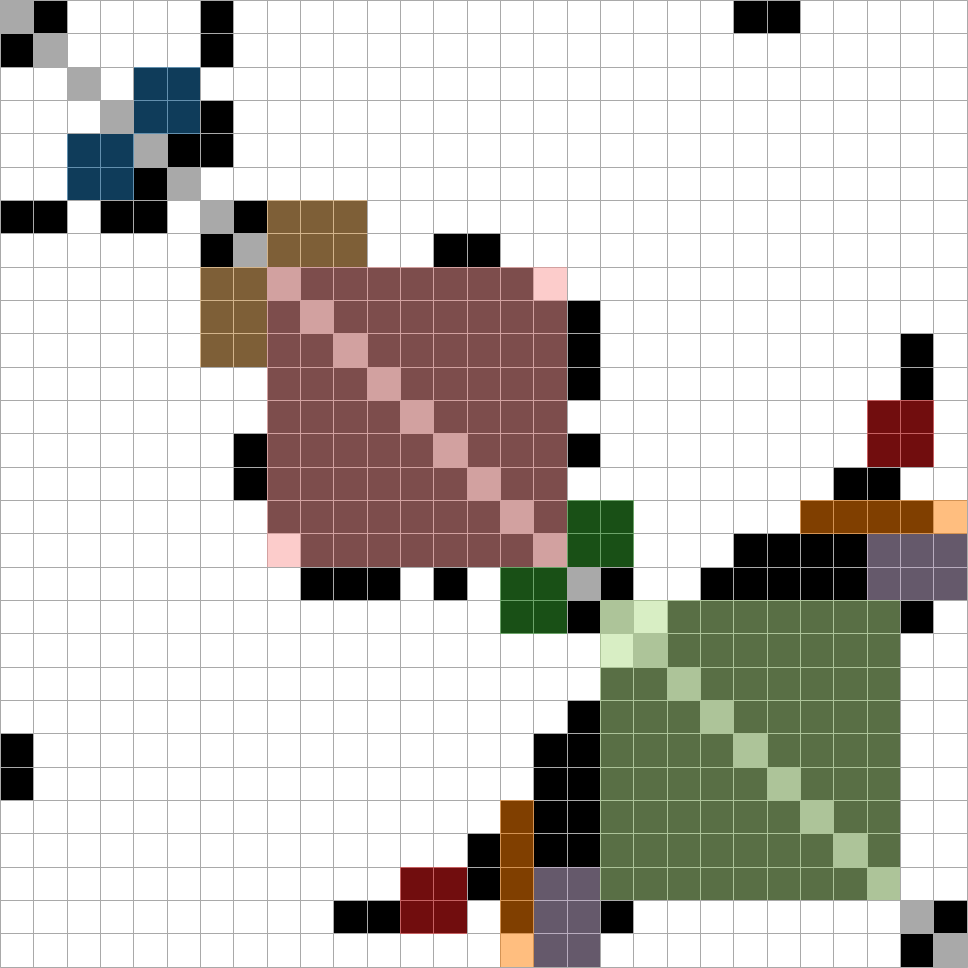} \\
        \raisebox{.08\linewidth}[0pt][0pt]{\rotatebox[origin=c]{90}{$\tau = 0.8$}} & \includegraphics[width=0.16\linewidth]{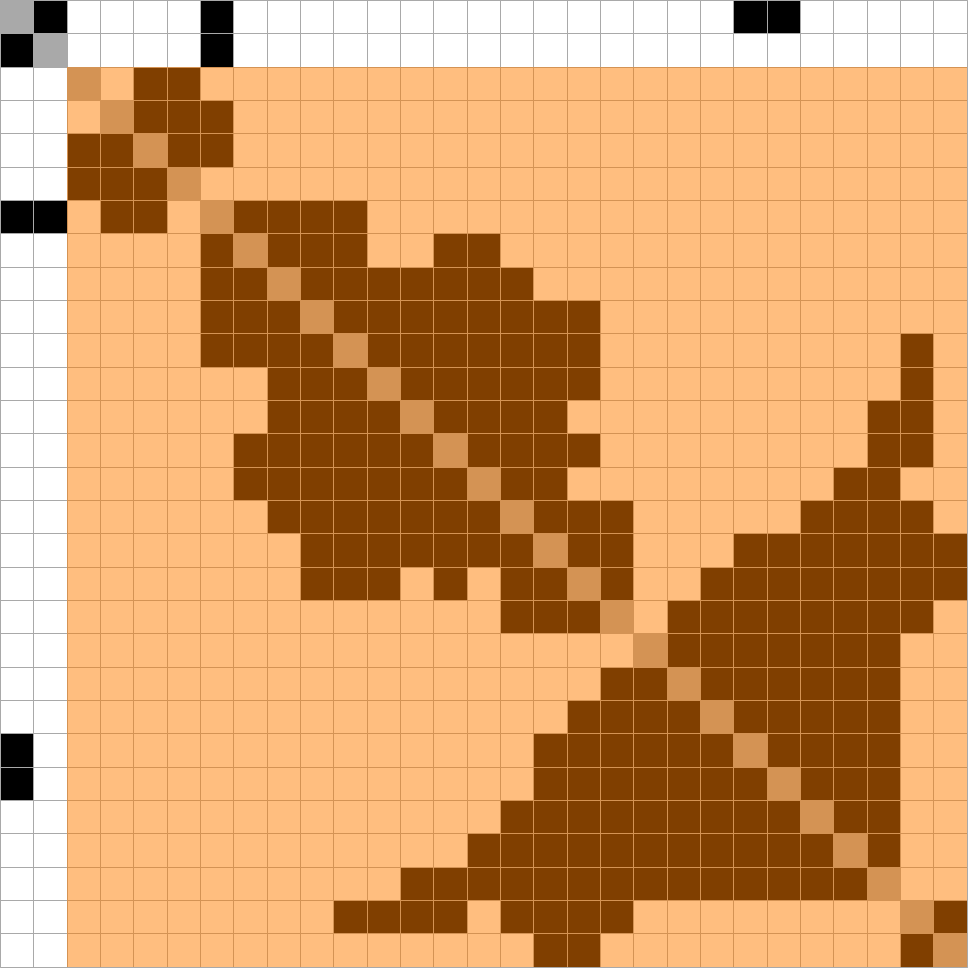} & \includegraphics[width=0.16\linewidth]{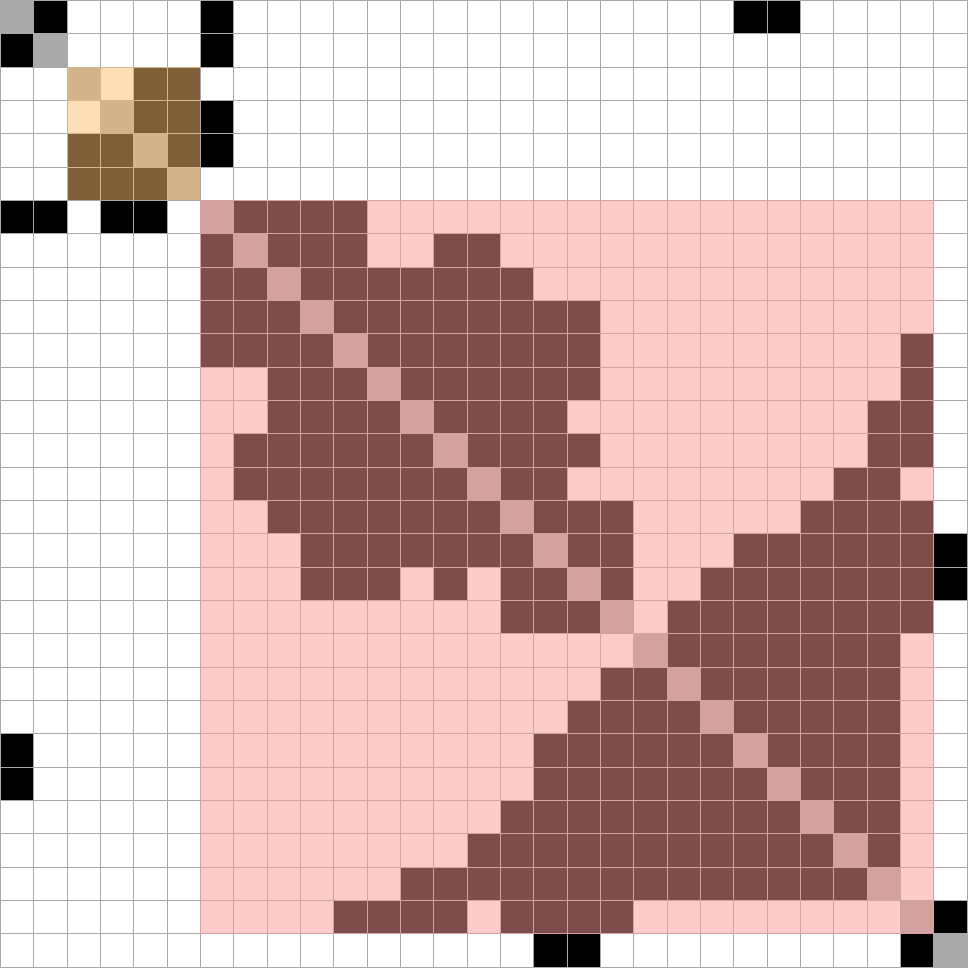} & \includegraphics[width=0.16\linewidth]{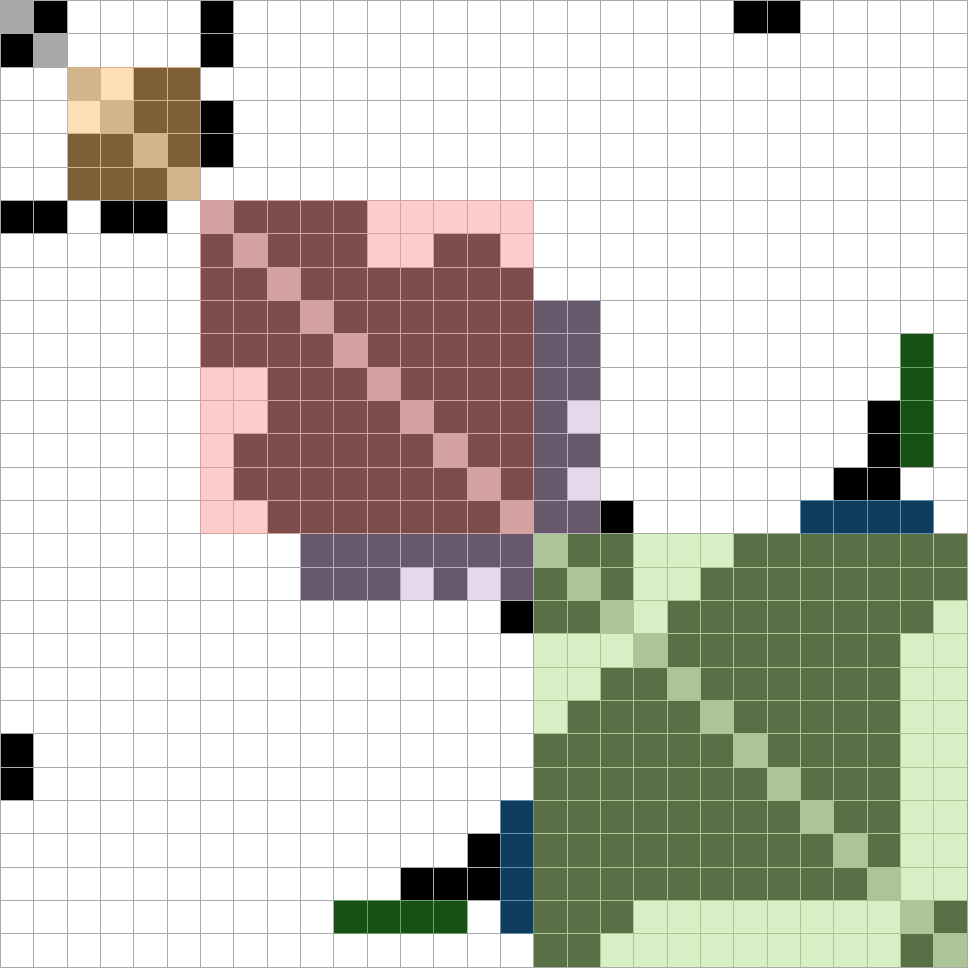} & \includegraphics[width=0.16\linewidth]{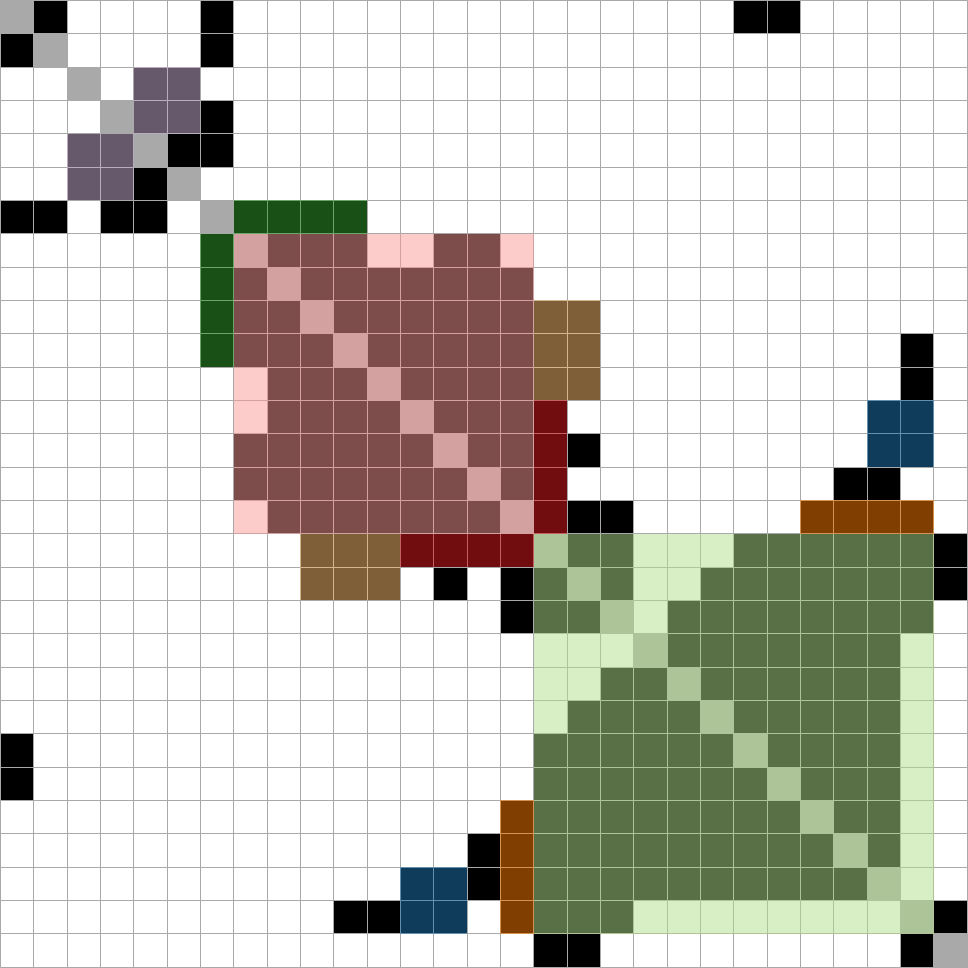} & \includegraphics[width=0.16\linewidth]{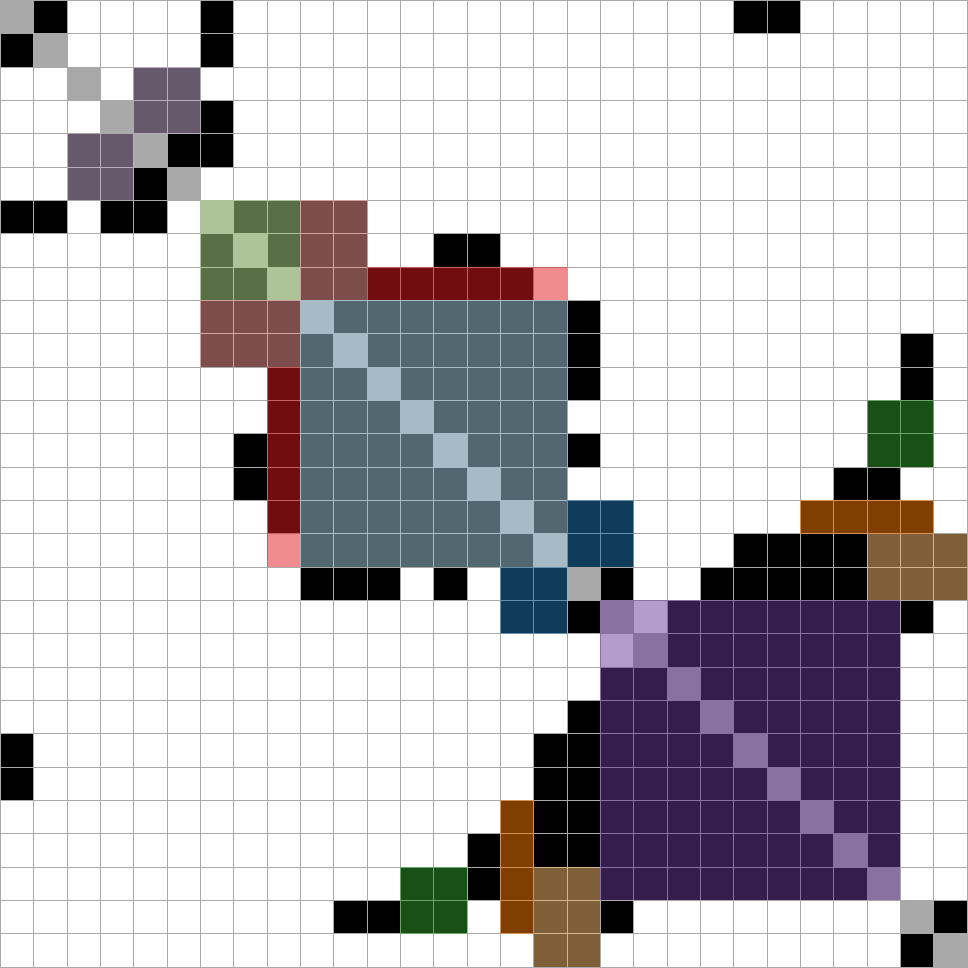} \\
        \raisebox{.08\linewidth}[0pt][0pt]{\rotatebox[origin=c]{90}{$\tau = 0.9$}} & \includegraphics[width=0.16\linewidth]{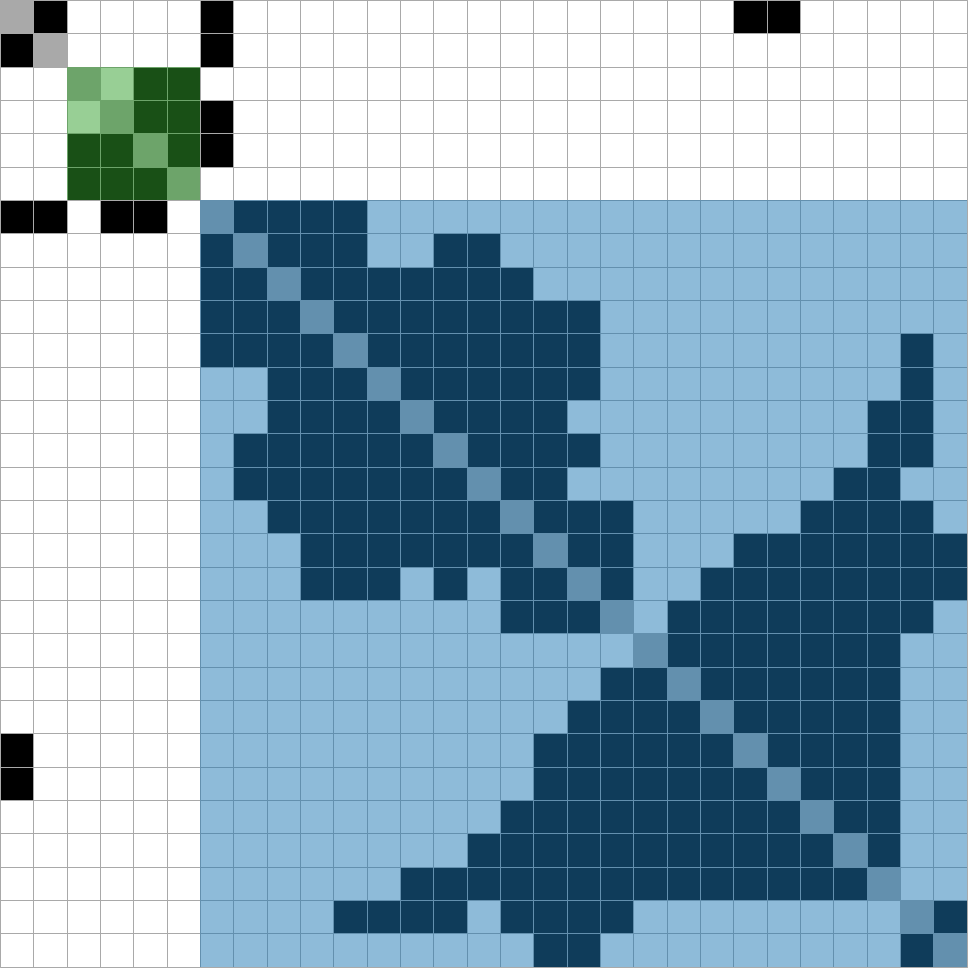} & \includegraphics[width=0.16\linewidth]{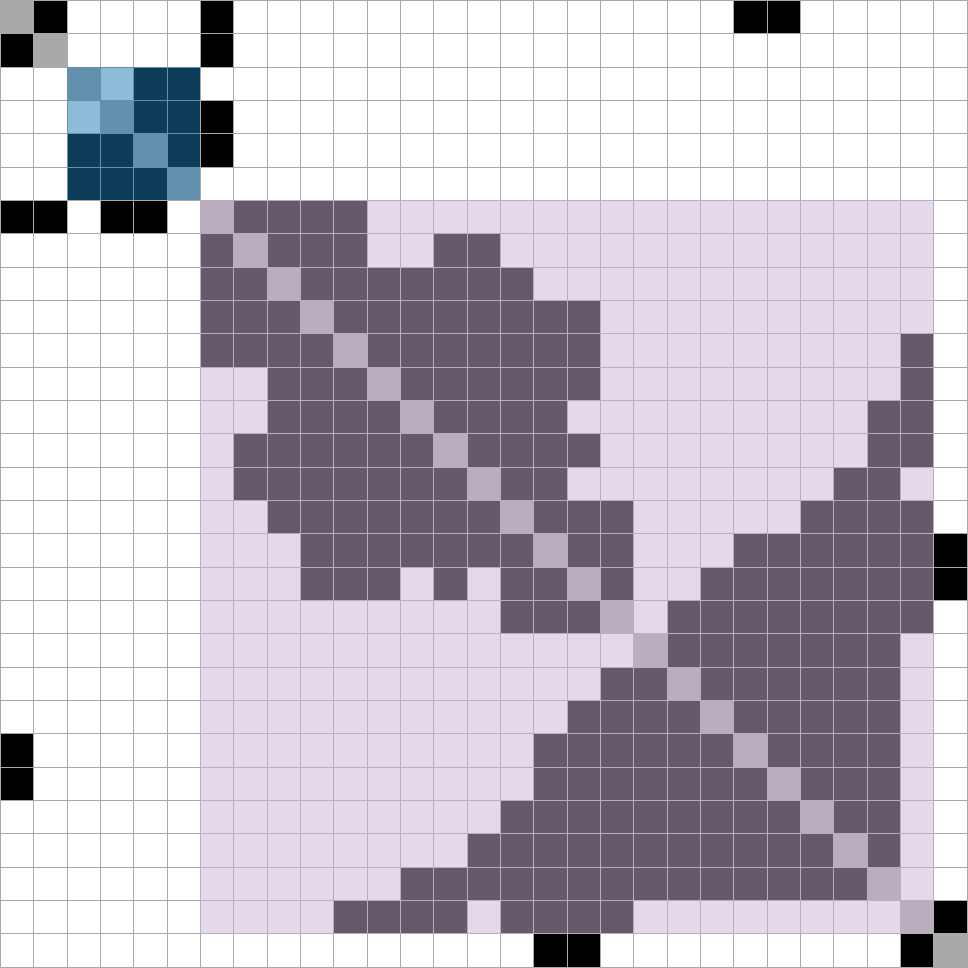} & \includegraphics[width=0.16\linewidth]{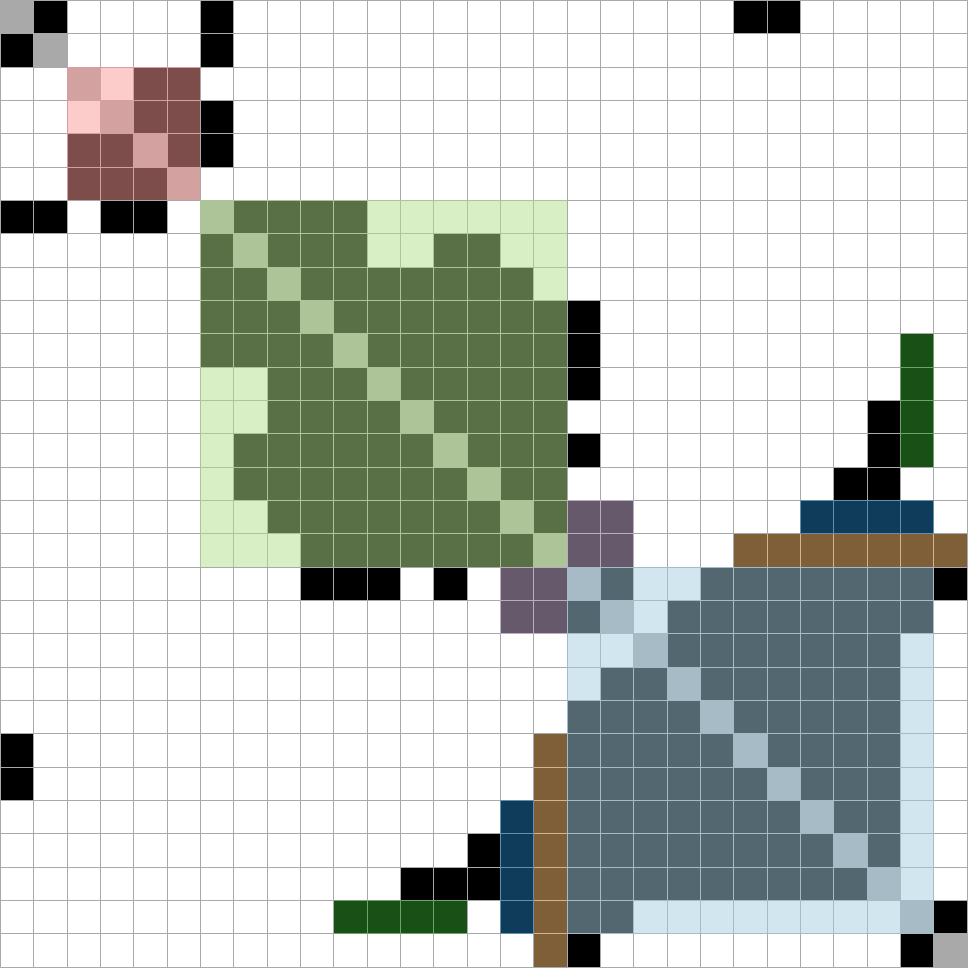} & \includegraphics[width=0.16\linewidth]{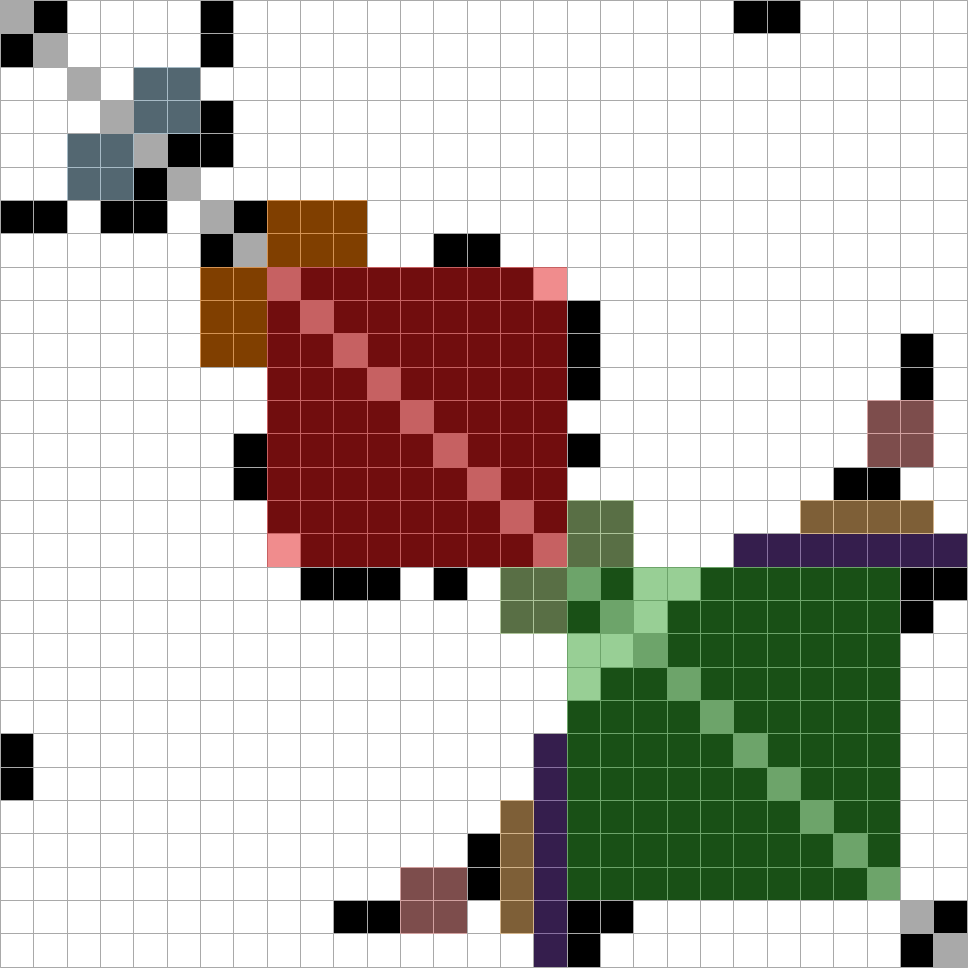} & \includegraphics[width=0.16\linewidth]{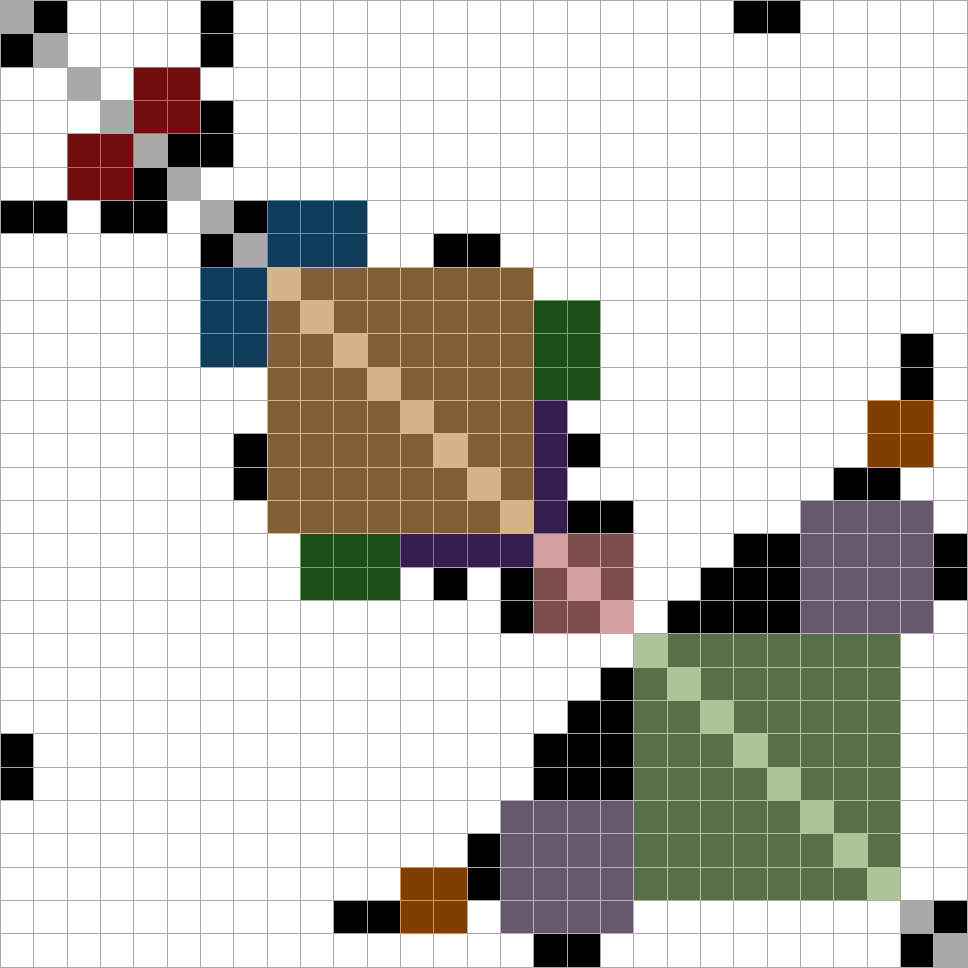} \\
        \raisebox{.08\linewidth}[0pt][0pt]{\rotatebox[origin=c]{90}{$\tau = 1.0$}} & \includegraphics[width=0.16\linewidth]{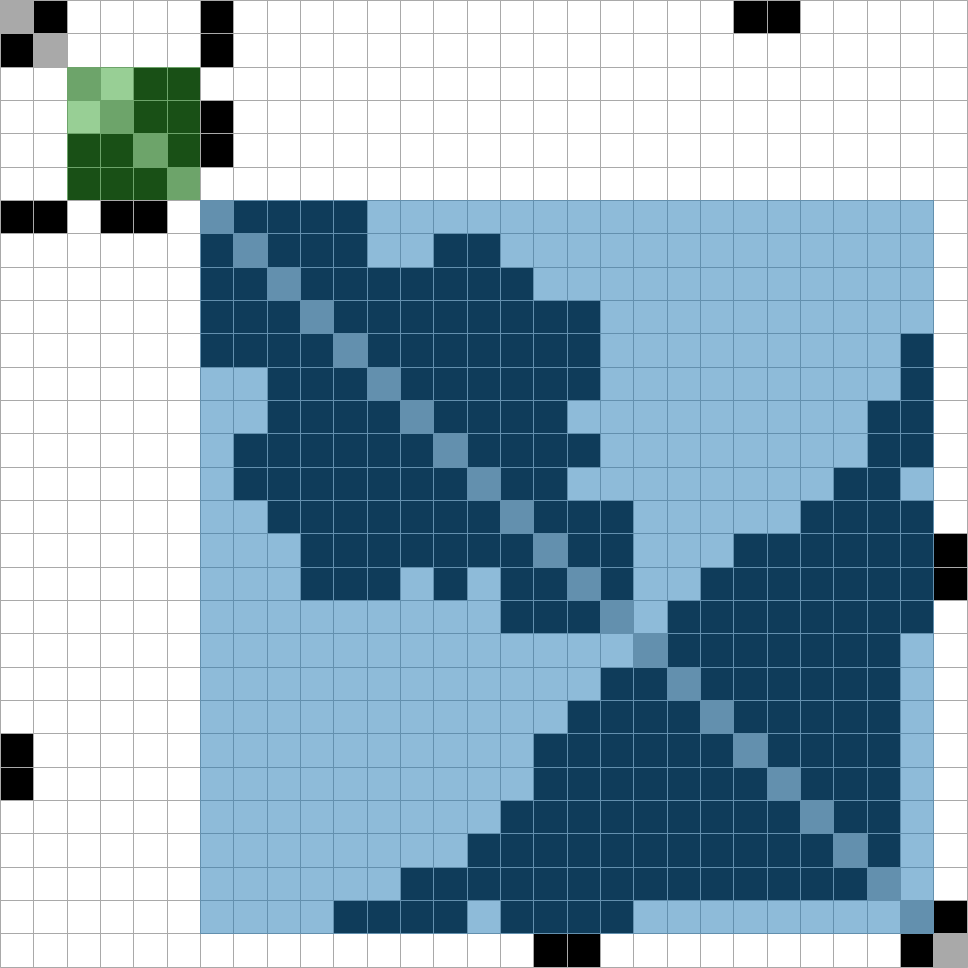} & \includegraphics[width=0.16\linewidth]{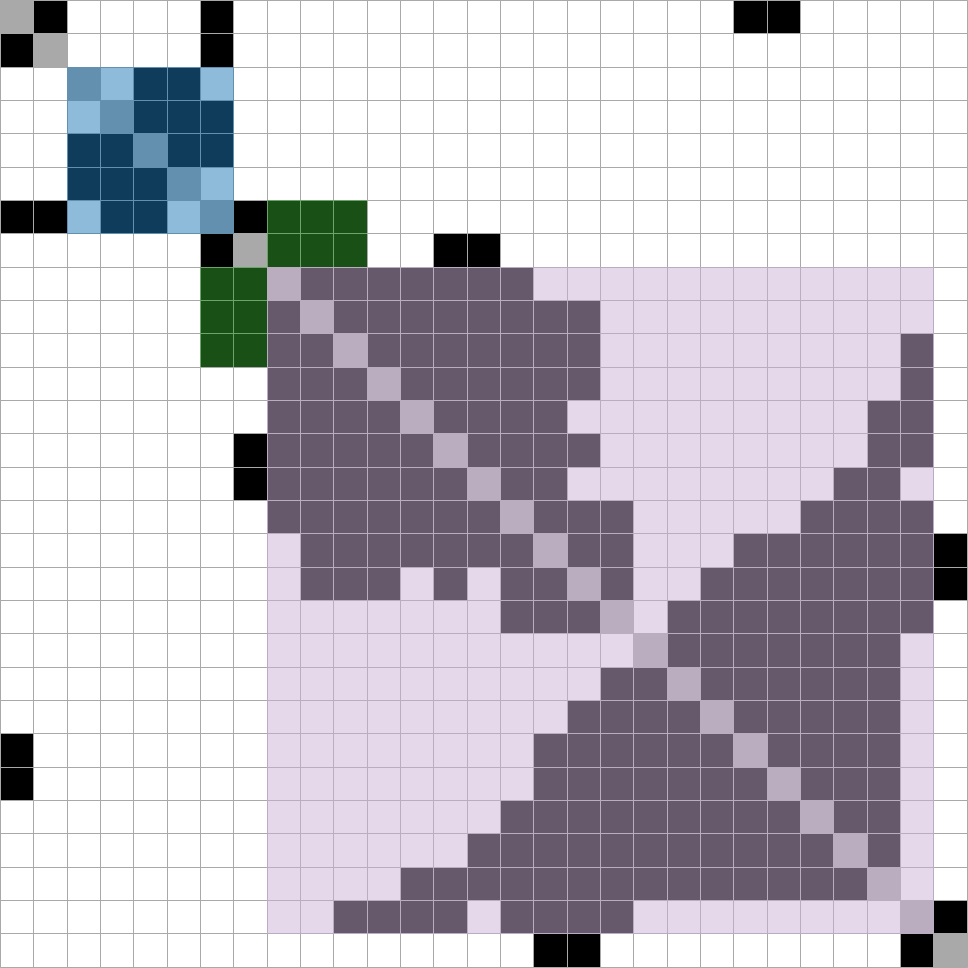} & \includegraphics[width=0.16\linewidth]{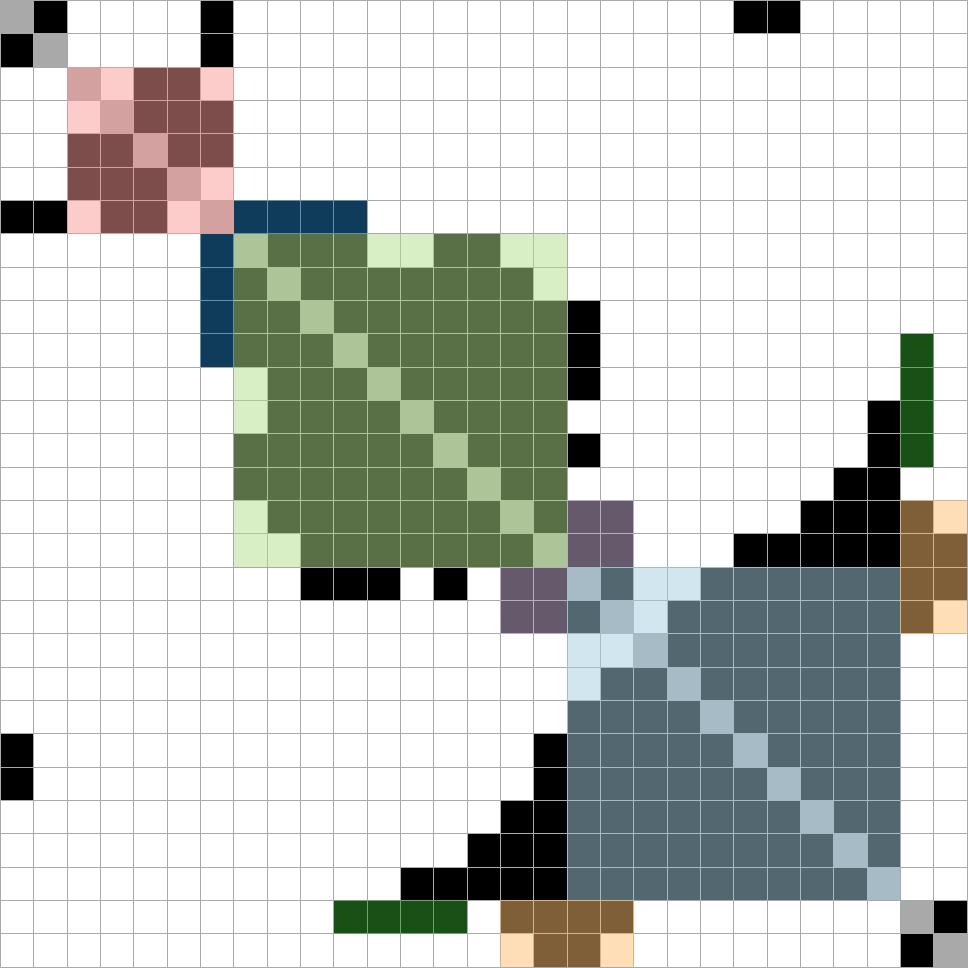} & \includegraphics[width=0.16\linewidth]{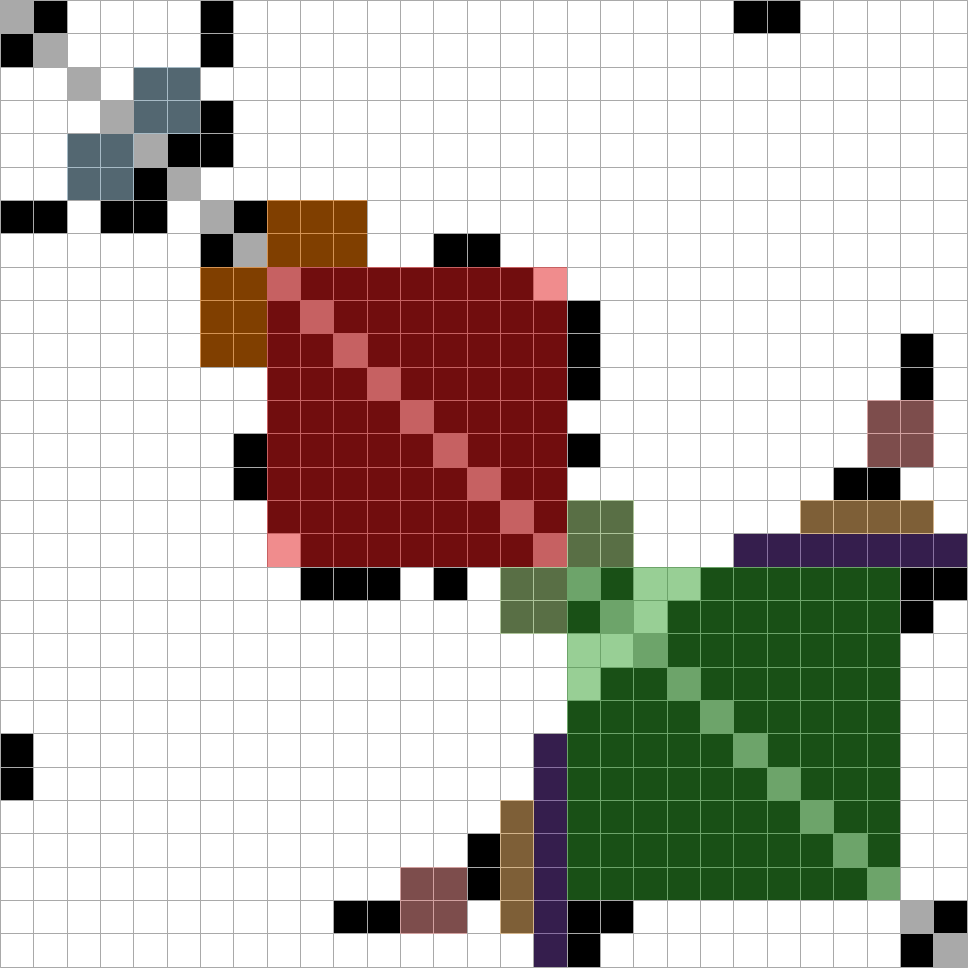} & \includegraphics[width=0.16\linewidth]{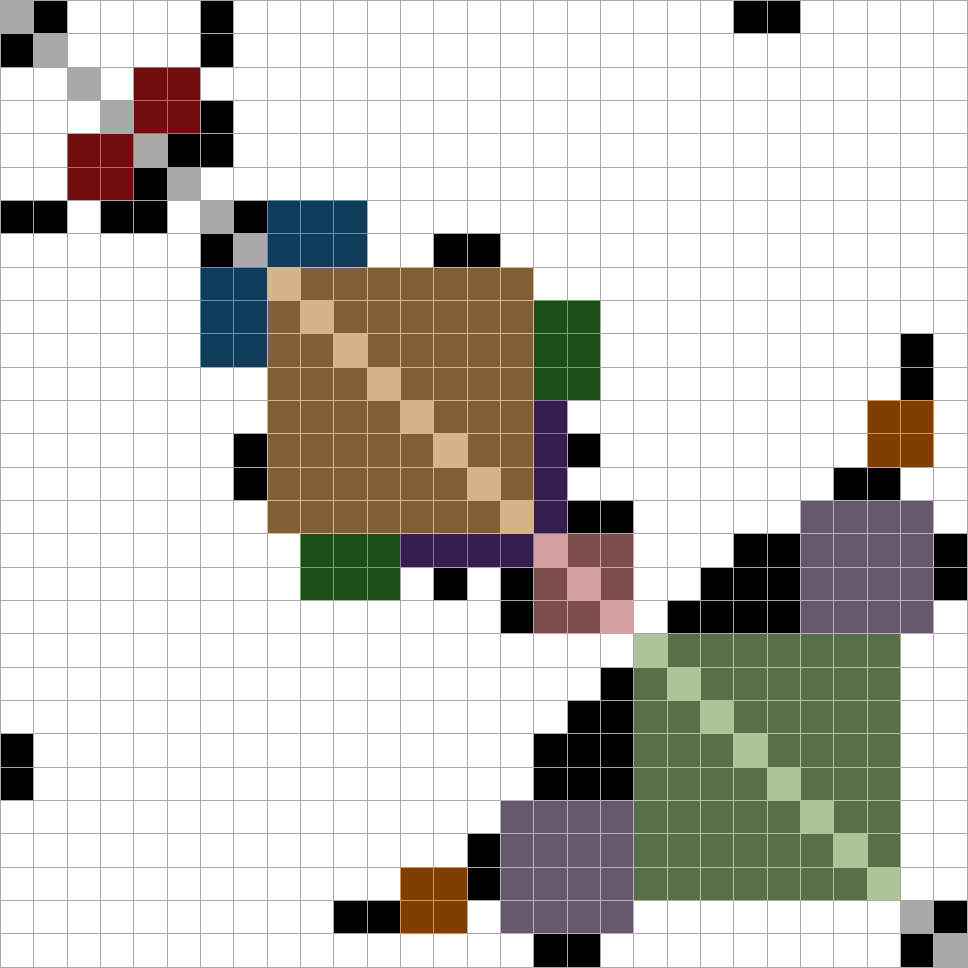}
    \end{tabular}
    \label{tab:parameter-experiment}
\end{table*}

\end{document}